%% file: LK2v1.3.tex
\documentclass[manuscript]{aastex61}
\usepackage{color}
\usepackage[normalem]{ulem}

\newcommand{\teff}{$T_{\rm eff}$}
\newcommand{\logg}{$\log g$}

\newcommand{\feh}{[Fe/H]}
\newcommand{\kepler}{{\it Kepler}}
\newcommand{\kms}{km\,s$^{-1}$}
\newcommand{\Ktwo}{{\it K}2}
\newcommand{\snrg}{S/N$_g$}

\received{-}
\revised{-}
\accepted{-}
\shorttitle{LAMOST Observations in 15 \Ktwo\ Campaigns: I}
\shortauthors{Jiangtao Wang et al.}
\begin{document}

\title{LAMOST Observations in 15 {\it K}2 Campaigns: I. Low resolution spectra from LAMOST DR6} %\footnote{This version fixes many bugs from v6.0 and introduces some new features, primarily in the way the author and affiliations are now marked up.}}

\correspondingauthor{J.-N. Fu, W. Zong}
\email{jnfu@bnu.edu.cn,weikai.zong@bnu.edu.cn}

\author{Jiangtao Wang}
\affil{Department of Astronomy, Beijing Normal University, 19 Avenue Xinjiekouwai, Beijing 100875, People's Republic of China}

\author{Jian-Ning Fu$^*$}
\affiliation{Department of Astronomy, Beijing Normal University, 19 Avenue Xinjiekouwai, Beijing 100875, People's Republic of China}

\author{Weikai Zong$^*$}
\affiliation{Department of Astronomy, Beijing Normal University, 19 Avenue Xinjiekouwai, Beijing 100875, People's Republic of China}

\author{M. C. Smith}
\affiliation{Shanghai Astronomical Observatory, Chinese Academy of Sciences, Shanghai 20030, People's Republic of China}

\author{Peter De Cat}
\affiliation{Royal Observatory of Belgium, Ringlaan 3, B-1180 Brussel, Belgium}

\author{Jianrong Shi}
\affiliation{Key Lab for Optical Astronomy, National Astronomical Observatories, Chinese Academy of Sciences, Beijing 100012, People's Republic of China}

\author{Ali Luo}
\affiliation{Key Lab for Optical Astronomy, National Astronomical Observatories, Chinese Academy of Sciences, Beijing 100012, People's Republic of China}

\author{Haotong Zhang}
\affiliation{Key Lab for Optical Astronomy, National Astronomical Observatories, Chinese Academy of Sciences, Beijing 100012, People's Republic of China}

\author{A. Frasca}
\affil{INAF--Osservatorio Astrofisico di Catania, Via S. Sofia 78, I-95123 Catania, Italy}

\author{C. J. Corbally}
\affil{Vatican Observatory Research Group, Steward Observatory, Tucson, AZ 85721-0065, USA}

\author{J. Molenda- \.{Z}akowicz}
\affil{Astronomical Institute of the University of Wroc{\l}aw, ul.\,Kopernika 11, 51-622 Wroc{\l}aw, Poland}

\author{G. Catanzaro}
\affil{INAF--Osservatorio Astrofisico di Catania, Via S. Sofia 78, I-95123 Catania, Italy}

\author{R. O. Gray}
\affil{Department of Physics and Astronomy, Appalachian State University, Boone, NC 28608, USA}

\author{Jiaxin Wang}
\affil{Department of Astronomy, Beijing Normal University, 19 Avenue Xinjiekouwai, Beijing 100875, People's Republic of China}

\author{Yang Pan}
\affil{Department of Astronomy, Beijing Normal University, 19 Avenue Xinjiekouwai, Beijing 100875, People's Republic of China}

%% Note that the \and command from previous versions of AASTeX is now
%% depreciated in this version as it is no longer necessary. AASTeX
%% automatically takes care of all commas and "and"s between authors names.

%% AASTeX 6.1 has the nePlease update your syste(Molenda-$\dot{Z}$akowicz et al. 2010b; McNamara et al. 2012; Huber et al. 2014, 2016m to include revtex4-1.clsw \collaboration and \nocollaboration commands to
%% provide the collaboration status of a group of authors. These commands
%% can be used either before or after the list of corresponding authors. The
%% argument for \collaboration is the collaboration identifier. Authors are
%% encouraged to surround collaboration identifiers with ()s. The
%% \nocollaboration command takes no argument and exists to indicate that
%% the nearby authors are not part of surrounding collaborations.

%% Mark off the abstract in the ``abstract'' environment.
\begin{abstract}
The LAMOST-{\sl K}2 (L{\sl K}2) project, initiated in 2015, aims to collect low-resolution spectra of targets in the {\sl K}2 campaigns, similar to LAMOST-{\sl Kepler} project. By the end of 2018, a total of 126 L{\sl K}2 plates had been observed by LAMOST. After cross-matching the catalog of the LAMOST data release 6 (DR6) 
with that of the {\sl K}2 approved targets, 
we found 160,619 usable spectra of 84,012 objects, most of which 
had been observed more than once. 
The effective temperature, surface gravity, metallicity, and radial velocity from 129,974 spectra for 70,895 objects are derived through the LAMOST Stellar Parameter Pipeline (LASP).
The internal uncertainties were estimated to be 
81 K, 0.15 dex, 0.09 dex and 5\,kms$^{-1}$, 
respectively,  when derived from a spectrum with a signal-to-noise ratio in the $g$ band (SNR$_g$) of 10. These estimates are based on results for targets with multiple visits. 
The external accuracies were assessed by comparing the parameters of targets in common with the APOGEE and GAIA surveys, for which we generally found linear relationships. A final calibration is provided, combining external and internal uncertainties for giants and dwarfs, separately. 
We foresee that these spectroscopic data will be used widely in different research fields, especially in combination with \Ktwo\ photometry.

\end{abstract}

%% Keywords should appear after the \end{abstract} command.
%% See the online documentation for the full list of available subject
%% keywords and the rules for their use.

%% From the front matter, we move on to the body of the paper.
%% Sections are demarcated by \section and \subsection, respectively.
%% Observe the use of the LaTeX \label
%% command after the \subsection to give a symbolic KEY to the
%% subsection for cross-referencing in a \ref command.
%% You can use LaTeX's \ref and \label commands to keep track of
%% cross-references to sections, equations, tables, and figures.
%% That way, if you change the order of any elements, LaTeX will
%% automatically renumber them.

%% We recommend that authors also use the natbib \citep
%% and \citet commands to identify citations.  The citations are
%% tied to the reference list via symbolic KEYs. The KEY corresponds
%% to the KEY in the \bibitem in the reference list below.

\section{Introduction} \label{sec:intro}
The \kepler\ spacecraft, launched by NASA in 2009 March, had as its main scientific 
goal the discovery of extrasolar Earth-like planets through transit events \citep{2010ApJ...713L..79K}. During its prime mission, \kepler\ collected unprecedented high-precision photometry for about 200,000 stars in a field of 115 square degrees between Cygnus and Lyrae \citep{2016RPPh...79c6901B}.
In 2014, the spacecraft shifted to observe the fields along the ecliptic plane due to pointing problem caused by failure of the second reaction wheel. 
The data produced by \Ktwo, like the prime \kepler~mission, were acquired in the short- and long-cadence modes, 
except that the time baseline was
reduced to approximately 80 days for each campaign \citep{2014PASP..126..398H}.

The \Ktwo\ mission collected
photometry for more than 400,000 stars during 20 campaigns (C0, C1,..., C19).
Those light curves are a treasure trove for many research areas, including exoplanets \citep{2015ApJ...809...25M}, asteroseismology \citep{2018ApJ...866..147C,2019MNRAS.489.4791S}, and eclipsing binaries \citep{2019MNRAS.487.4230S}.
Nevertheless, for many applications, an in-depth exploitation of these data requires the 
knowledge of precise atmospheric parameters.
For instance, 
optimal seismic models are more reliable and easier to find when 
the effective temperature (\teff), surface gravity (\logg) and metallicity (\feh) have been determined from spectroscopic measurements beforehand \citep{2011A&A...530A...3C,2018Natur.554...73G}. 
Unfortunately, the Ecliptic Plane Input Catalog \citep[EPIC;][]{2016ApJS..224....2H} of the \Ktwo\ sources provides atmospheric parameters derived from multi-band photometry, which do not have a high enough accuracy for the demands of asteroseismology. 
Therefore, to fully exploit the \Ktwo\ data, many follow-up programs have been initiated.
This includes spectroscopic ones, such as the Mauna Kea Spectroscopic Explorer \citep[MSE;][]{2019arXiv190303157B} and Twinkle \citep{2019LPI....50.1388J} (similar to the \kepler\ follow-up programs APOGEE \citep{2017AJ....154...94M, 2017ApJS..233...23S, 2018ApJS..239...32P}), the California-Kepler Survey \citep[CKS;][]{2017AJ....154..107P}, and the K2-HERMES Survey \citep{2018AJ....155...84W}, as well as photometric ones like the SkyMapper \citep{2019MNRAS.482.2770C}.

Based on the experience gained during previous observing campaigns, the Large Sky Area Multi-Object Fiber Spectroscopic Telescope (LAMOST, aka, Gou
Shou Jing Telescope) has proved to be an ideal instrument for follow-up spectroscopic observations on targets within the \kepler\ field \citep[LAMOST-\kepler\ project,][]{2015ApJS..220...19D, 2020arXiv200810776F, 2018ApJS..238...30Z}. After two rounds of observations from 2012 to 2017, the LAMOST-\kepler\ project collected more than 220,000 spectra of 156,390 stars, providing useful parameters for exoplanet statistics \citep{2016PNAS..11311431X, 2018PNAS..115..266D, Muldersetal2016}, precise asteroseismology \citep{2014A&A...564A..27D} and stellar activity \citep{2016A&A...594A..39F,Karoffetal2016,2017ApJ...849...36Y}.

One of the biggest obstacles in carrying out the LAMOST-\kepler\ project was the fact that the \kepler\ field is observed mainly during the summer season, when the nights available at the Xinglong Observatory are reduced due to the monsoons and the instrument maintenance. 
Unlike \kepler, the \Ktwo\ mission has a much wider sky coverage, consisting of 20 fields 
identical in size to the \kepler\ field, 
uniformly distributed along the ecliptic. 
This has given more opportunities to observe the \Ktwo\ fields with LAMOST, excluding only those with a declination lower than $-10$~degrees
that are not observable.
As a consequence, this has enlarged the research areas of interest from asteroseismology,
stellar activity and exoplanet discovery to gravitational lensing,  AGN variability, and supernovae \citep{2014PASP..126..398H}.
The LAMOST-\Ktwo\ (L\Ktwo) project, initiated in 2015, 
aims to collect spectra for as many EPIC stars as possible, with the final 
goal of producing a very large, homogeneous catalog of atmospheric parameters for stars of various 
types and in different evolutionary stages, from the pre-main sequence phase to evolved objects like white dwarfs.
Moreover, during its regular survey phase, LAMOST had already collected spectra for targets within several \Ktwo\ campaigns before the
L\Ktwo-project began. This is very valuable for the study of, for example, pulsating stars and binaries.

In this paper we summarize the main results gained from the analysis of spectra of \Ktwo\ targets from the L\Ktwo project and the sixth LAMOST
data release (DR6).
The paper is organized as follows. In Sect.~\ref{sec:observation}, we present the observations and the step we have made
toward the completion of the  L\Ktwo\ program.
Section~\ref{sec:SPD} 
describes the library of spectra obtained within the L\Ktwo\ project and during the regular LAMOST survey in the \Ktwo{} fields.
%the selection of the \Ktwo{} objects for observations of the L\Ktwo\ project and the identification of \Ktwo-field objects already observed by LAMOST until DR6.}
In Sect.~\ref{subsec:para}, we present the atmospheric parameters for the L\Ktwo\ stars, and discuss their uncertainties and systematics. We propose external and internal calibrations to homogenize these data with those of other spectroscopic surveys.
In Sect.~\ref{sec:distri}, we discuss interesting objects identified on the basis of 
their stellar parameters. We give a final summary in Section~\ref{subsec:disc}.

\section{Observations} \label{sec:observation}

\begin{figure*}
\includegraphics[width=18cm]{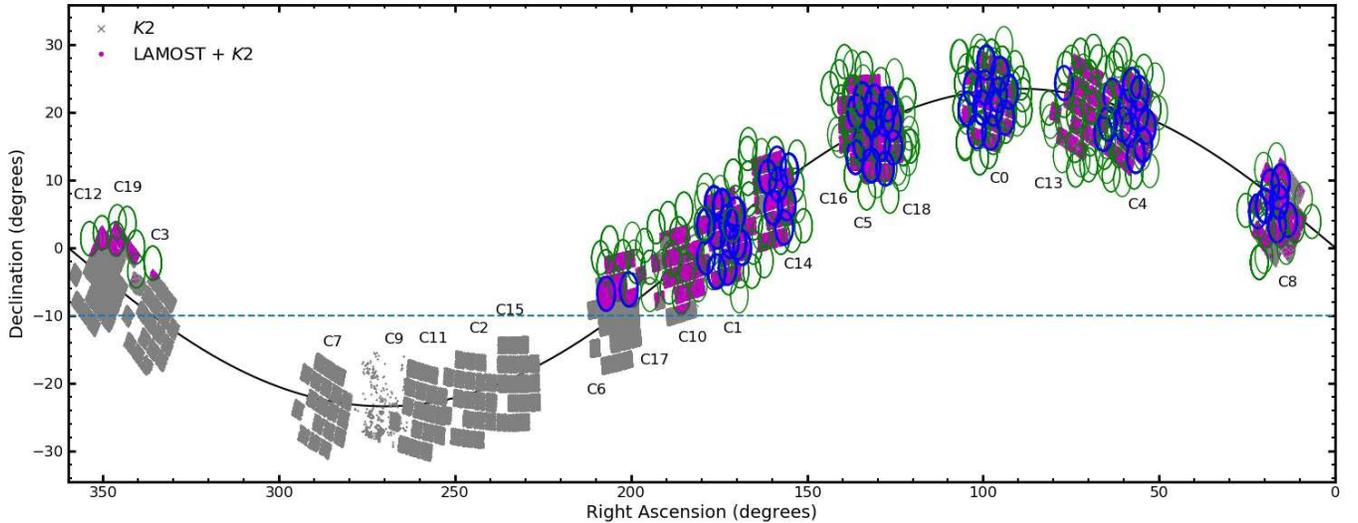}
\centering
\caption{Sky distribution of \Ktwo\ targets and LAMOST DR6 cross-matched sources in the 20 campaigns
of the \Ktwo\ mission.
The gray symbols denote the stars with \Ktwo\ photometry and the magenta dots refer to the \Ktwo\ targets observed with LAMOST. The area coverage of the L\Ktwo\ project and the LAMOST general survey in the \Ktwo\ fields is indicated by blue and green circles, respectively. The cyan dashed line at $-10\degr$ indicates the declination limit for LAMOST observations. The black solid line indicates the ecliptic plane.
\label{fig:sky1}}
\end{figure*}

LAMOST, equipped with 4000 fibers on its focal plane, is capable to simultaneously collect spectra for about 3600 targets, with a few hundreds of fibers pointing to the sky.
In order to improve the efficiency of these observations, each footprint is advised to contain targets covering a certain range of magnitude.
This leads to four types of LAMOST plates, namely, very bright (V), bright (B), medium-brightness (M) and faint (F) plates, respectively, according to the target
brightness range \citep[see details in][]{2012RAA....12.1243L, 2015RAA....15.1095L}. 
The L\Ktwo\ plates are typically V- and B-plates 
because \Ktwo\ photometry has been collected for stars brighter than 16th magnitude.
However, because the fields along the ecliptic plane, as observed for the L\Ktwo\ project, are not crowded, fainter objects needed to be added to fill the fibers.
Similar to the L{\sl K}-project, each of the 20 \Ktwo\ campaigns is divided into 14 circular fields where the central position is determined by a bright central star ($V < 8$). 
We note that the \Ktwo\ fields include a few unobserved regions corresponding to failed CCD modules on board of \kepler. 
No LAMOST plate was assigned in these positions.
The plates within the L\Ktwo-project have a nomenclature of 'KP'+'RA'+'DEC'+'Plate type' where 'KP' denotes the plates belong to the projects related to follow-up observation of {\sl Kepler/K}2 targets. There has been a revision of nomenclature after 2017 October, i.e. 'KP' has been replaced by 'KII' for the L\Ktwo-project, in order to distinguish
them from those of the L{\sl K} project.

The first L\Ktwo\ plate was exposed in 2015 December, and a total of 126 plates has been collected until 2018 January.
We acquired 1, 84, 31, and 10 plates during 1, 50, 20, and 9 nights in 2015, 2016, 2017, and 2018, respectively.
We have given a higher priority to bright plates,
i.e. V- and B-plates with exposures of 3$\times$600\,s and 3$\times$1500\,s, respectively.
Additionally, there are 6 M-plates observed with an exposure time of 3$\times$1800\,s each. 
The total shutter open time is of the order of 100 hours for all the L\Ktwo\ plates, without taking overhead into account.

LAMOST performs a general regular survey of as many as possible targets across the entire northern hemisphere with declination higher than $-10$ degrees \citep[see, e.g.,][]{2012RAA....12.1243L}. 
Therefore, it is likely that several plates show overlap with some of the \Ktwo\ campaigns, especially for C0 and C13 where the density of the stellar sources in the regular survey is high. We have also collected the spectra of these common targets in our library,  
based on the criteria in section \ref{sec:cross}.
However, most of them have only a few targets in common.
As a consequence, 401 of these 652 plates have a number less than 100 targets with \Ktwo\ photometry.

Figure\,\ref{fig:sky1} shows the sky coverage of all stars observed by LAMOST until DR6 stamped over the \Ktwo\ photometric targets.
We clearly see that the L\Ktwo\ plates were observed over the campaigns with right ascension lower than 210 degrees, namely
{C0, C1, C4, C5, C6, C8, C10, C13, C14, C16, C17, and C18}. We note that all these L\Ktwo\ stamped campaigns have stars in common with the LAMOST regular survey and \Ktwo\ photometry. Another three campaigns, C3, C12 and C19, are also found with common stars.

\section{Spectra library}
\label{sec:SPD}

The spectral library of the present paper contains spectra of
\Ktwo\ sources coming from both the L\Ktwo-project and other sub-projects of the regular LAMOST survey.
All these spectra can be downloaded from the LAMOST DR6\footnote{http://dr6.lamost.org/} website, which contains about nine million low-resolution spectra. The calibrated spectra were produced through the 2.7.5 version of the LAMOST 2D and 1D pipelines \citep[see][for details]{2015RAA....15.1095L}.

\subsection{Cross-identification} \label{sec:cross}
Unlike the \kepler\ observations that were mainly focused on asteroseismology and exoplanets \citep{2013ApJ...767...95D}, \Ktwo\ was approved to cover a wider
range of astrophysical topics, including gravitational lensing \citep{2013ApJ...779L..28G}, asteroids and comets \citep{2017A&A...599A..44S}, and AGNs
\citep{2013ApJ...766...16E}.
However, we selected only stellar sources observed by the \Ktwo\ mission as targets
for our LAMOST low-resolution spectroscopic observations.
There are 15 out of 20 \Ktwo\ campaigns with a declination higher than $-10$ degree that could be observed with LAMOST (see Figure \ref{fig:sky1}).
They include 306\,838 out of 406\,270 objects 
collected with \Ktwo\ photometry.

The cross-match of \Ktwo\ and LAMOST DR6 catalogs was made with TOPCAT \citep{2005ASPC..347...29T}, based on a criterion of
distance separation less than 3.7 arcsecs, which is a bit larger than the 3.0 arcsecs of the LAMOST-\kepler\ project \citep{2018ApJS..238...30Z}.
This value for the search radius was adopted because the diameter of the fiber is 3.3 arcsecs and the pointing precision is 0.4 arcsecs.
We note that the position of the fibers was offset for stars brighter than $V=11^{\rm m}$ to prevent saturation during exposure, but this was taken into account.
Besides, we only selected spectra with signal-to-noise ratio in the SDSS $g$ band \snrg$\geq6$, which are mentioned as ``qualified spectra'' in the present paper.

The cross-match produced a final catalog that includes 
160\,629 low-resolution qualified spectra of 84\,012 \Ktwo\ objects.
This amounts to 
27.38\% of all the observable \Ktwo\ stars, or 20.68\% of 
%the entire \Ktwo\ \color{blue}{observed} objects.}
all the stars with \Ktwo\ observations.
Table\,\ref{tab:cross} reports information on the observed plates, the number of sources cross-matched with the \Ktwo\ catalog, and the number of sources with derived parameters.
It also includes the number of sources with multiple visits. In total, more than 30,000 sources were observed more than once.
The sky position of these objects is depicted in Fig.\,\ref{fig:sky1}, which shows that almost all the \Ktwo\ fields with DEC $>-10\degr$ were observed with LAMOST.
We note that C3, C12 and C19 can only be observed in summer time, i.e. during the monsoon season. In that period, the observing time for LAMOST is heavily reduced and
the telescope is often closed for maintenance. This explains why we have very few data in these fields, as apparent from Fig.\,\ref{fig:sky1}.

\begin{figure}
\centering
\includegraphics[width=8cm]{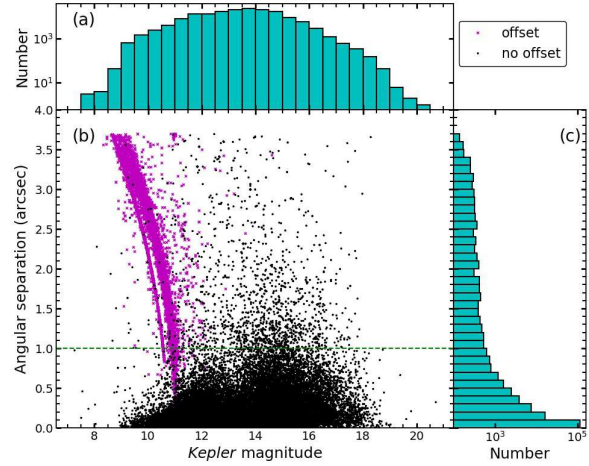}
\caption{Angular separation between \Ktwo\ and LAMOST cross-matched sources versus their \kepler\ magnitude 
(panel b) and the projected histograms (a and c).
The magenta dots denote the bright sources for which the position of the LAMOST fiber was purposely offset,
otherwise they are displayed as black dots.
The green dashed line marks an angular separation of 1.0 arcsec.
\label{fig:cross}}
\end{figure}

Figure\,\ref{fig:cross} contains all the L\Ktwo\ targets from LAMOST DR6 cross-matched with the \Ktwo\ catalog.  Figure\,\ref{fig:cross}(b) shows the distribution of the angular separation between the coordinates of the LAMOST DR6 and \Ktwo\ catalogs as a function of the {\sl K}p magnitude.
The higher the angular separation, the more doubtful is the cross-identification. 
This distribution is projected to one dimension of angular separation (Figure\,\ref{fig:cross}(c)) and {\sl K}p (Figure\,\ref{fig:cross}(a)), respectively. We find that most of the cross-matched objects ($\sim94.36\%$) display an angular separation in the range of 0--1 arcsecs. 
Nevertheless, there is quite a high fraction of bright objects, with brightness in the range of 9$^m$--11$^m$, whose input coordinates have larger shifts in RA and DEC.
Those targets, flagged with ``offset" from DR6 catalog, should be treated with caution. Some of these have been purposely shifted to prevent saturation. If they were removed, the proportion of ``large" angular separation will be reduced to a fraction of $\sim3.6\%$.
In general, our cross-identified catalog contains objects with brightness mainly in
the range of 10$^m$--18$^m$,
with the majority found around 12$^m$-16$^m$.
This is different from the L{\sl K}-project where there is a sharp decrease in the number of targets with brightness fainter than 14th magnitude \citep{2018ApJS..238...30Z}.  
In the the L\Ktwo\ project, the collected plates include not only V- but also some B- and M-plates.
It changes the faint tail of the distribution of the magnitudes of the observed targets when compared to the LAMOST-\kepler\ project.

\startlongtable
\begin{deluxetable*}{crrrr}
\tablecaption{General information of the common stars between \Ktwo\ and LAMOST DR6 catalogs from 2011 to 2018.
The bottom lines give the summary of the observations of the L\Ktwo-project where we give 
%{\color{cyan} \sout{the number of different cross-matched objects that have been observed (Unique) and}}
the number of targets that have been observed one (1$\times$), two (2$\times$), three (3$\times$), four (4$\times$), and at least five (+5$\times$) times.
\label{tab:cross}}
%\tablehead{\colhead{Year} & \colhead{L\Ktwo\ Plate} & \colhead{Other Plate}  & \colhead{Spectra} & \colhead{Parameter} }
\tablehead{\colhead{Year} & \colhead{L\Ktwo\ Plate} & \colhead{Survey Plate}  & \colhead{Spectra} & \colhead{Parameter} }
\startdata
2011 & -  & 15   & 481   & 312 \\
2012 & -  & 91   & 15735 & 11874 \\
2013 & -  & 123  & 19648 & 16616 \\
2014 & -  & 130  & 21635 & 18849 \\
2015 & 1  & 93   & 18003 & 15673 \\
2016 & 84 & 80   & 47515 & 38758 \\
2017 & 31 & 70   & 23129 & 17037 \\
2018 & 10 & 50   & 14473 & 10855 \\
\hline
Total      & 126  & 652 & 160619 & 129974  \\
%Unique     &      &     &    &   \\
\hline
%\enddata
Visits &  &   & Sources & Parameter \\
%\startdata
1$\times$     &      &     & 48280  & 41634 \\
2$\times$  &      &     & 20877  & 17445 \\
3$\times$  &      &     & 8392   & 6827  \\
4$\times$  &      &     & 3404   & 2753 \\
+5$\times$ &      &     & 3059   & 2236 \\
\enddata
\end{deluxetable*}

Table \ref{tab:spec} lists the catalog of the LAMOST low-resolution spectra collected for objects with \Ktwo\ photometry.
In the present paper we print only the first 3 lines as an example. The full table can be downloaded at LAMOST DR6 value added catalogs website,
which contains the following columns:
\\
(1) Obsid: a unique identification (ID) of the calibrated spectrum; 
\\ 
(2) EPIC: the cross-identified ID from the EPIC catalog where a coordinate separation of 3.7 arcsec is used as the limit (the nearest star is chosen if more than one star is identified);
\\
(3) RA (2000): the input right ascension (epoch J2000.0) to which the fiber was pointed (in hh:mm:ss.ss); 
\\
(4) Dec. (2000): the input declination (epoch J2000.0) to which the fiber was pointed (in dd:mm:ss.ss);
\\
(5) \snrg: the signal-to-noise ratio (S/N) of the spectrum in SDSS g band, which is an indicator for the quality of the spectrum;
\\
(6) {\sl K}p: the \kepler\ magnitude from  \kepler\ Data Search\footnote{http://archive.stsci.edu/k2/data\_search/search.php} website.
\\
(7) SpT: the spectral type of the target, calculated by the LAMOST 1D pipeline;
\\
(8) UTC: the Coordinated Universal Time at mid-exposure (in yyyymmddThh:mm:ss);
\\
(9) C: the number of the \Ktwo\ campaign to which the object belongs;
\\
(10) $\Delta$d: angular separation between the equatorial coordinates of the LAMOST fiber and cross-identified source in the EPIC catalog (in arcsec);
\\
(11) Filename: the name of the corresponding LAMOST 1D fits file.

\startlongtable
\begin{longrotatetable}
\begin{deluxetable*}{llccccrcccl}
\tablecaption{The spectral database of L\Ktwo\ sample obtained by cross-identification of LAMOST DR6 and \Ktwo\ catalogs.\label{tab:spec}}
\tablehead{
\colhead{Obid} &
\colhead{EPIC} &
\colhead{RA} &
\colhead{Dec} &
\colhead{\snrg} &
\colhead{{\sl K}p} &
\colhead{SpT} &
\colhead{UTC} &
\colhead{C} &
\colhead{$\Delta$d} &
\colhead{Filename} \\
\colhead{} &
\colhead{} &
\colhead{(hh:mm:ss.ss)} &
\colhead{(dd:mm:ss.ss)} &
\colhead{} &
\colhead{(mag)} &
\colhead{} &
\colhead{(yyyymmddThh:mm:ss)} &
\colhead{} &
\colhead{(arcsec)} &
\colhead{}
}
\startdata
\input{Table02.tex}
\enddata
\tablecomments{The table has a total of 160\,619 entries which can be obtained through the link {\bf abcdefgh} until we are ready to submit the paper.}
\end{deluxetable*}
\end{longrotatetable}

\subsection{Spectra quality distribution} \label{subsec:chara}
\begin{figure*}
\centering
\includegraphics[width=15cm]{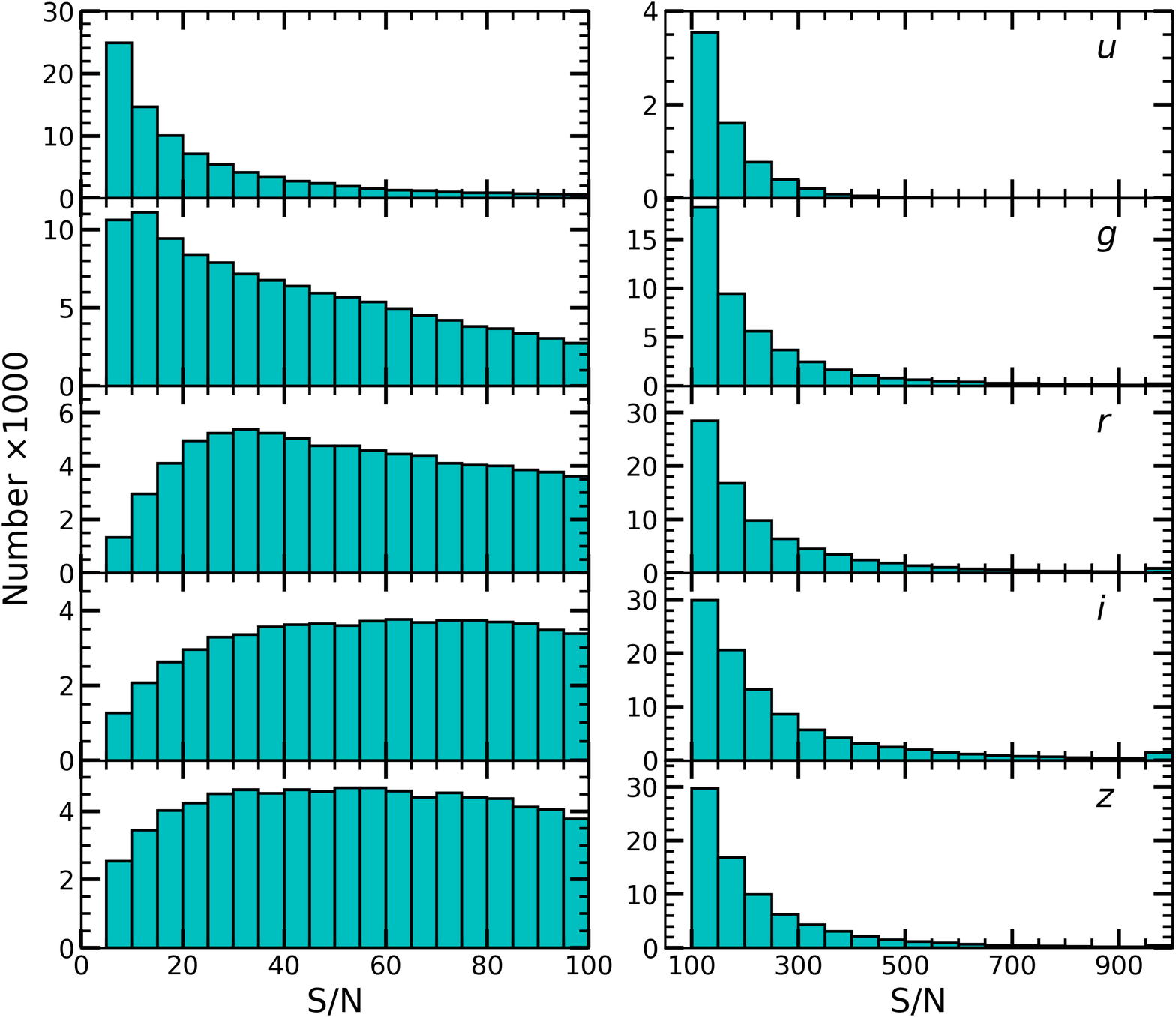}
\caption{Distributions of the S/N in the SDSS u, g, r, i, and z bands (from top to bottom) for the spectra of LAMOST-\Ktwo\ common stars. The left and right panels show the S/N ranges 6--100 and 100--1000, with bin sizes 5 and 50, respectively. 
\label{snr}}
\end{figure*}

A sensitive indicator of the quality of these spectra is their S/N, typically in the SDSS $g$ band (\snrg).
Figure\,\ref{snr} presents the distribution of the S/N of these spectra in the SDSS $u$, $g$, $r$, $i$, and $z$ bands. The left panel shows S/N distributions in the range 6--100 with
a bin size of 5, 
while a ten times wider range (S/N\,=\,100--1000) is shown in the right panel with a bin size of 100. 
Similar to what we did in the L{\sl K}-project, a spectrum is investigated by the LASP analysis pipeline for the determination of atmospheric parameters if its \snrg\ is larger than 6
if it was obtained in a dark night, or larger than 15 
for observations from other nights \citep[see, e.g.,][]{2015RAA....15.1095L,2018ApJS..238...30Z}. The entire catalog contains {149\,996, 138\,880, and 86\,949 spectra with \snrg\,$\geq 10$, \snrg\,$\geq 15$ and \snrg\,$\geq 50$ corresponding to a fraction of 93.39\%, 86.47\% and 54.13\%, respectively}. This shows that LAMOST has produced a very high percentage of spectra with a good quality for the objects in \Ktwo\ campaigns.

\section{stellar parameters} \label{subsec:para}

The standard LASP pipeline was applied to the spectra of our library to derive the atmospheric parameters and radial velocities if their \snrg\ was higher than the threshold values, essentially depending on spectral type. 
LASP incorporates two modes, the Correlation Function Interpolation \citep[CFI,][]{2012SPIE.8451E..37D} method and the Universit\'e de Lyon Spectroscopic analysis Software \citep[ULyss,][]{2009A&A...501.1269K, 2011A&A...525A..71W} to determine stellar parameters. 
In practice, the spectral type from the LAMOST 1D pipeline is used for the first evaluation of the input spectrum.
If the star is too hot or too cold (spectral type before late-A and after K), then the atmospheric parameters are not determined by LASP.
For input spectra of stars with a spectral type of late-A, F, G, or K, CFI is applied to obtain initial values of \teff, \logg, and \feh\ in the following way. First, \teff\ of the input spectrum is determined by comparison with synthetic spectra calculated for a grid of values of \teff. The resulting \teff-value is fixed before searching for the optimal solution in the parameter space of \logg\ and \feh. The values with the highest reliability given by CFI are used as initial input parameters for the application of ULySS. The final LASP parameters are those giving the smallest squared difference between the observation and the model \citep[see details in][]{2015RAA....15.1095L}.

A final number of 129,974
atmospheric parameters for 70,895 stars were produced with the LASP pipeline (v.2.9.7), which corresponds to a fraction of 81\,\% and 84\,\% for spectra and objects, respectively. We now count $\sim28$\,\% of the observable \Ktwo\ targets with homogeneously derived parameters with the same instrument and derived from the same pipeline. 
This is currently the largest homogeneous catalog of spectroscopically derived parameters for those targets.
To properly use these data, it is necessary to do a quality control, i.e. an evaluation of the precision and accuracy of the derived parameters.
To this aim we made both "internal" tests, essentially based on the objects with multiple observations, and "external" checks based on the comparison with parameters from the literature coming from other large spectroscopic surveys. The atmospheric parameters of some variable stars vary significantly during the period, such as eclipsing binaries, RR Lyrae stars,
and other variable stars listed by \citet{2017ARep...61...80S} and \citet{2016MNRAS.456.2260A}. Those stars have been flagged in Table \ref{tab:aps} in this paper. 
Finally, 92,853 spectra of 53,421 non-variable targets with LAMOST stellar parameters were used for the external comparison, while the internal uncertainty is estimated from 21,118 stars observed twice or more.
Our data of the L\Ktwo\ sample are contained in Table\,\ref{tab:aps}, which reports the entire catalog of parameters derived with the LASP pipeline. 
It contains the following columns:
\\
(1)--(4): Same as Table\,\ref{tab:spec},
\\
(5) \teff:  the effective temperature (in K);
\\
(6) \logg: the surface gravity (in dex);
\\
(7) \feh: the metallicity (in dex);
\\
(8) RV: the heliocentric radial velocity (in \kms);
\\
 (9) Comment: Special star candidate labels, including metal-poor stars (MPs), very metal-poor stars (VMPs),  high-velocity stars (HVs; cf.  Section~\ref{sec:distri} for details) and the types of variable star.
\\
All the uncertainties are provided by the LASP pipeline.

\startlongtable
\begin{longrotatetable}
\begin{deluxetable*}{cccccccccc}
\tablecaption{The L\Ktwo\ stellar parameter database obtained through the LASP pipeline. \label{tab:aps}}
\tablehead{
\colhead{Obsid} &
\colhead{EPIC} &
\colhead{RA} &
\colhead{Dec} &
\colhead{\teff} &
\colhead{\logg} &
\colhead{\feh} &
\colhead{RV} &
\colhead{Comment}\\
\colhead{} &
\colhead{} &
\colhead{(hh:mm:ss.ss)} &
\colhead{(dd:mm:ss.ss)} &
\colhead{(K)} &
\colhead{(dex)} &
\colhead{(dex)} &
\colhead{(\kms)} &
\colhead{}
}
\startdata
\input{Table03.tex}
\enddata
\tablecomments{The table has a total of 129\,974 lines, only a part is given here, the complete table is available on the data release website.}
\end{deluxetable*}
\end{longrotatetable}

\subsection{Internal uncertainties} \label{sec:In-calibration}

As mentioned above, LAMOST observed a fraction of targets two or more times (see Table\,\ref{tab:cross}). Those objects are useful to estimate the
internal uncertainties of the atmospheric parameters and RV, if we treat the values coming from different spectra of the same star as fully independent
measurements randomly distributed around the mean.
An unbiased way to estimate the internal errors is based on the following equations:
\begin{equation}
\Delta P_{i} = \sqrt{n/(n-1)} (P_{i} - \overline{P}),
\label{eq:deltap}
\end{equation}
{\noindent where $i \in [0,n]$ and $n$ represent the $i$th set of values of the parameter $P_{i}$ and the total number of measurements for the same star, respectively.}
$\overline{P}$ denotes the average value for a given object.
Equation\,\ref{eq:deltap} was applied to each parameter for obtaining the unbiased internal uncertainties of $\Delta$\teff, $\Delta$\logg, $\Delta$\feh, and $\Delta$RV.

\begin{figure}%[ht!]
\centering
\includegraphics[width=8cm]{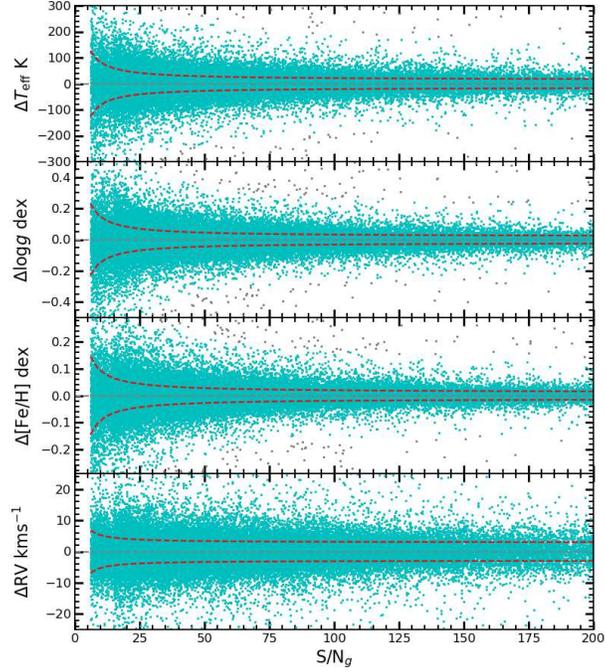}
\caption{Scatter plots of the deviation from the mean as a function of the signal-to-noise ratio \snrg\ for \teff, \logg, \feh, and RVs (from top to bottom) for the L\Ktwo\ targets with multiple visits (dots). The dashed lines represent the best fits to these data with a function of the type
of Eq.\,\ref{eq:fit}. The gray points indicate that the outliers in the bin are beyond 6$\sigma$, and the cyan points are used for fitting. \label{err}}
\end{figure}

Figure\,\ref{err} shows the deviations of each parameter from the  average as a function of \snrg. As expected, the deviation clearly decreases with 
increasing \snrg\ for all parameters up to \snrg\,$\sim 100$. 
For larger \snrg\ values, the scatter seems to attain an almost constant value for all the derived quantities.
We use a reciprocal function to fit those parameters and RV, in the range of \snrg\,$\in [6, 200]$, as
\begin{equation}
|\Delta P| = a_1 \centerdot {\rm S/N}_g^{-1} + b_1,
\label{eq:fit}
\end{equation}
{\noindent with a bin size of \snrg\,$= 10$. The coefficients, $a_1$ and $b_1$, of the best fit for each parameter are given in Table\,\ref{tab:incalib}.}
According to these fitting curves, the internal errors of \teff, \logg, \feh, and RV are 81\,K, 0.0.15\,dex, 0.09\,dex and 5\,\kms\ when \snrg\,$= 10$,
respectively. For \snrg\,$\geq 50$, the curves tend to nearly constant uncertainties of about 28\,K, 0.05\,dex, 0.03\,dex and 3\,\kms\ for \teff, \logg, \feh, and RV, respectively.

\begin{deluxetable}{lrrrr}
\tablecaption{The coefficients of the optimal fittings for each parameters and RV.\label{tab:incalib}}
\tablewidth{600pt}
\tablehead{
\colhead{} &
\colhead{\teff} &
\colhead{\logg} &
\colhead{\feh} &
\colhead{RV}
}
\startdata
$a_1$ & 668.38 & 1.25  & 0.79  & 23.37 \\
$b_1$ & 14.53   & 0.02 & 0.01 & 2.89 \\
\enddata
\end{deluxetable}

\subsection{External accuracy} \label{sec:Ex-calibration}
A comparison with results of other spectroscopic surveys can help to estimate the accuracy of our parameter determination.
The atmospheric parameters for \Ktwo\ campaigns C0--C8 were collected from a variety of catalogs by \citet{2016ApJS..224....2H}.
However, \Ktwo\ finally  released 20 campaigns before it retired.
We found two large surveys that contains suitable volume to compare with our data,  namely the APOGEE \citep[Apache Point Observatory Galactic Evolution Experiment: SDSS DR16][]{2020AJ....160..120J} and GAIA DR2  \citep{2018A&A...616A...1G}
For our stars with multiple observations, we adopted the parameters derived from the LAMOST spectrum with the largest \snrg. 
We divide non-variable targets into two samples by a sharp cut at $\log{g}_{\rm LAMOST} = 3.5$ dex, where stars with $\log{g}_{\rm LAMOST} < 3.5$ dex are classified as giants and the others as dwarfs.
In their recent paper, \citet{2020arXiv200906843Z} show that the $T_\mathrm{eff}$ values from Gaia, which typical uncertainty is 300\,K, become doubtful for stars with high extinction ($A_G > 0.8$). Therefore, in our external comparisons we include only the RV values measured by Gaia.
We found that 4,017 and 51,259 targets 
are overlapping
with the L\Ktwo\ non-variable target in APOGEE and GAIA, respectively, within 3.7 arcsec errors. 
We selected targets with \snrg\,$>$\,15 to do  a reliable comparison. Finally,  we found 1,307 giant and 2,519 dwarf stars in common with APOGEE, and 20,020 stars in common with GAIA DR2.

\begin{figure*}
\centering
\includegraphics[width=6cm]{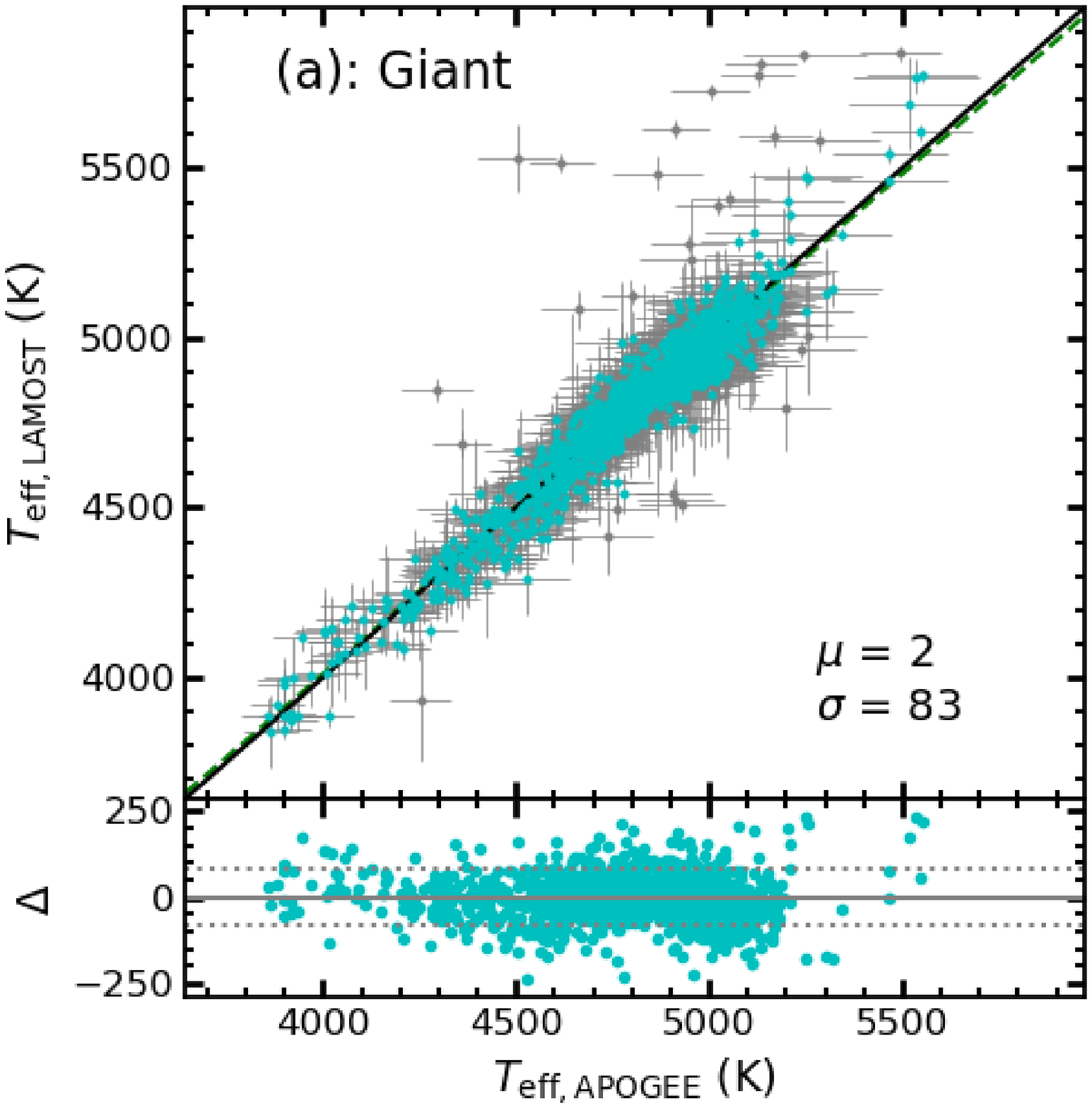}\includegraphics[width=6cm]{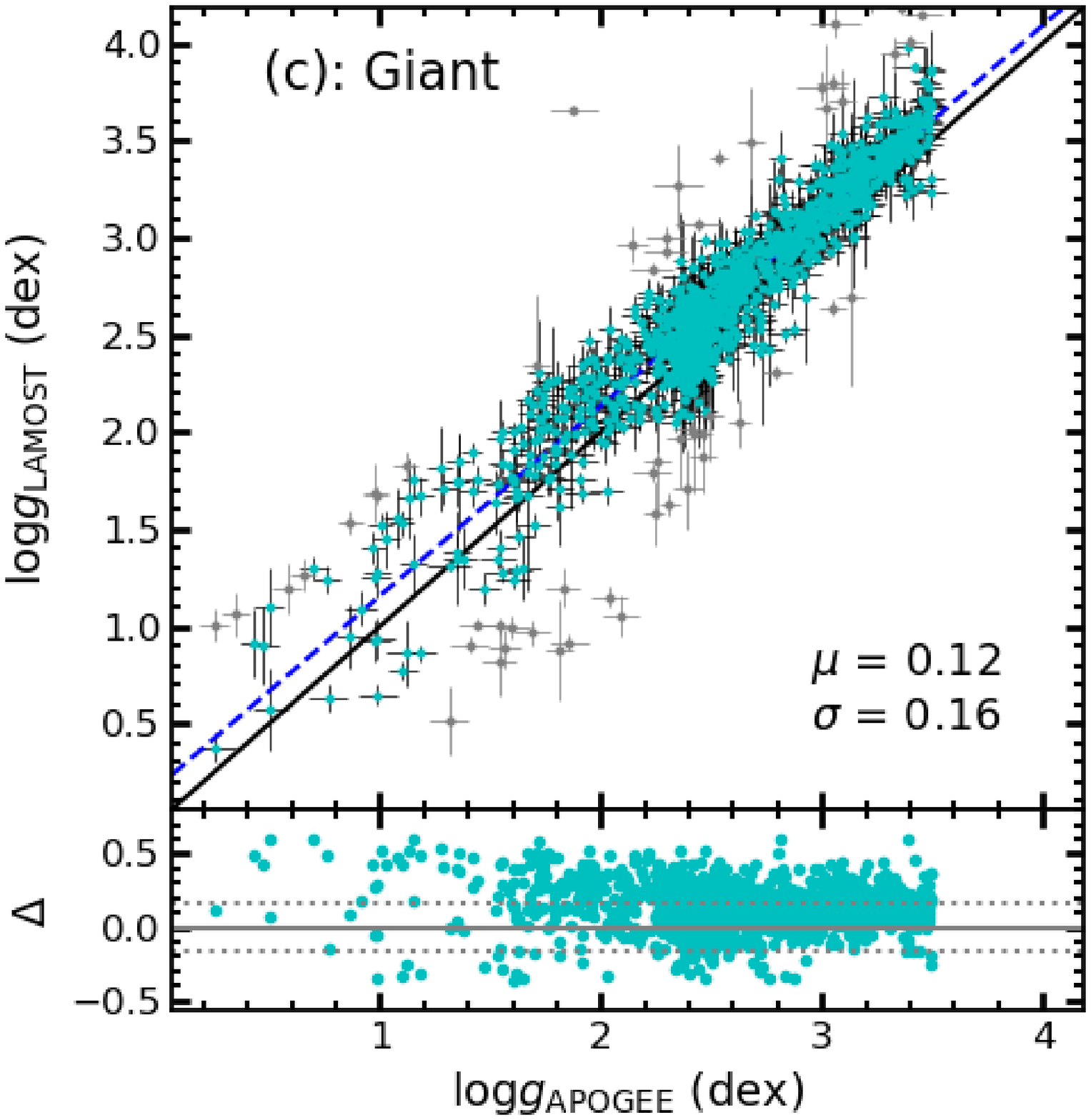}\includegraphics[width=6cm]{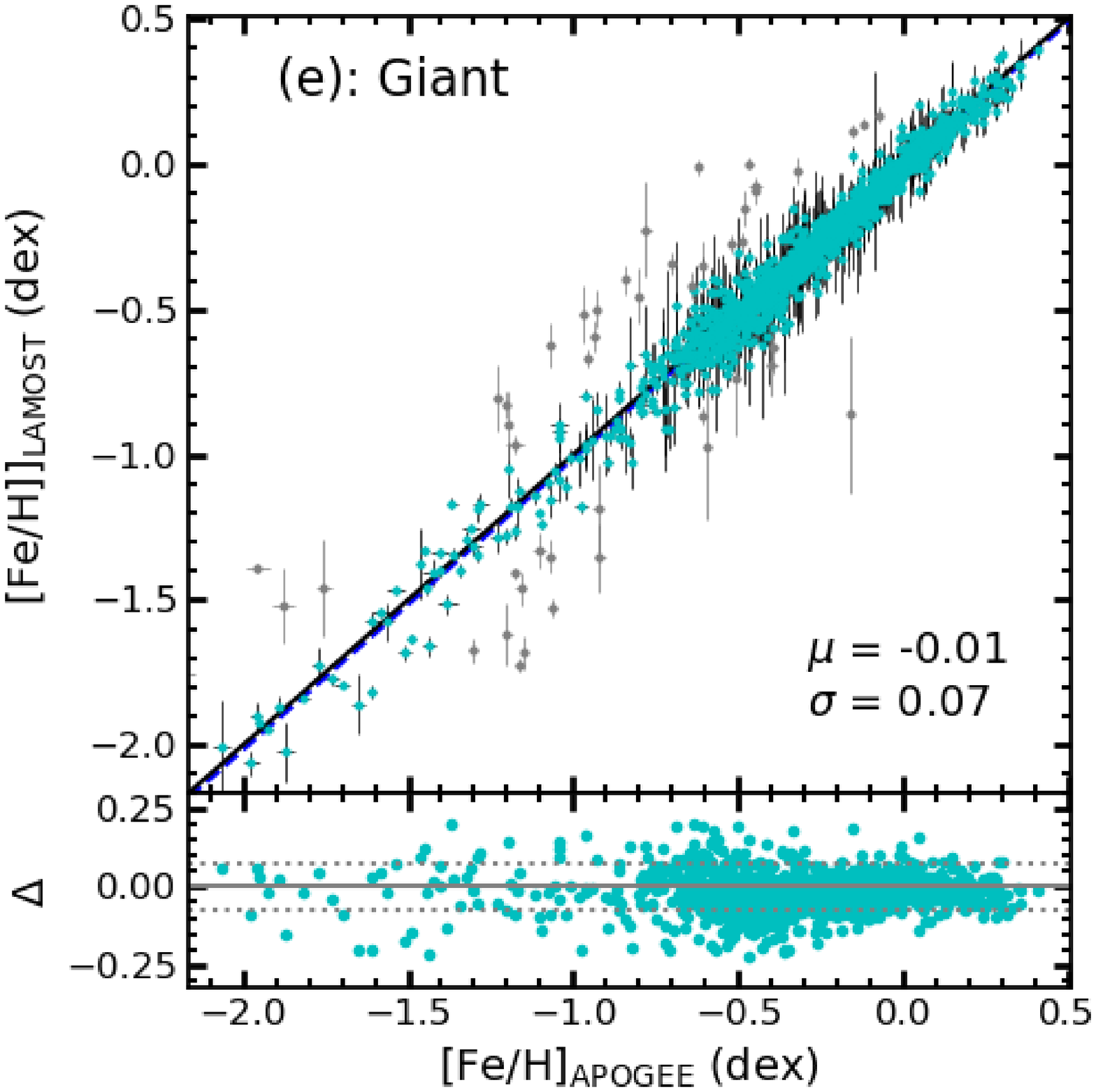}
\includegraphics[width=6cm]{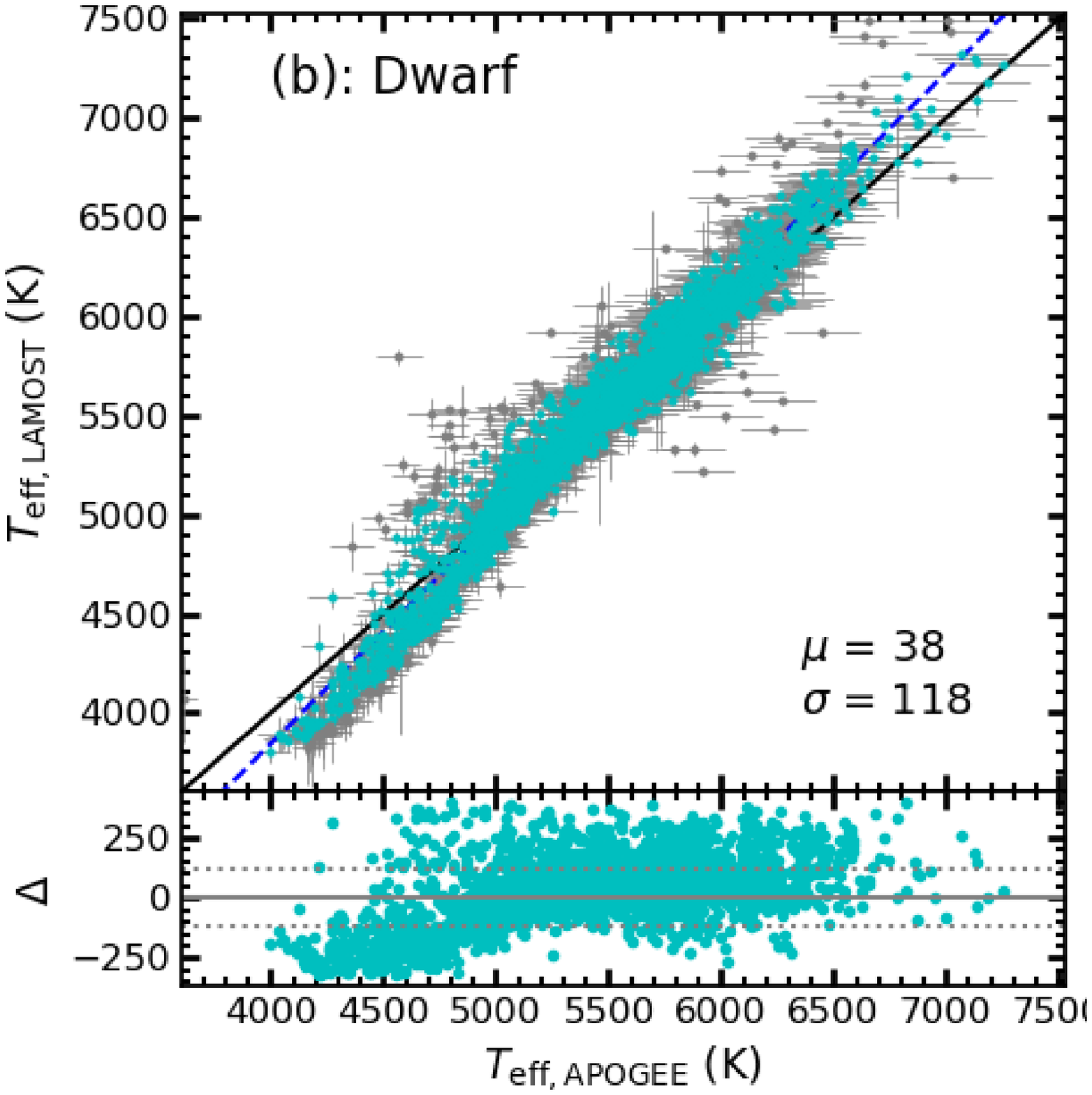}\includegraphics[width=6cm]{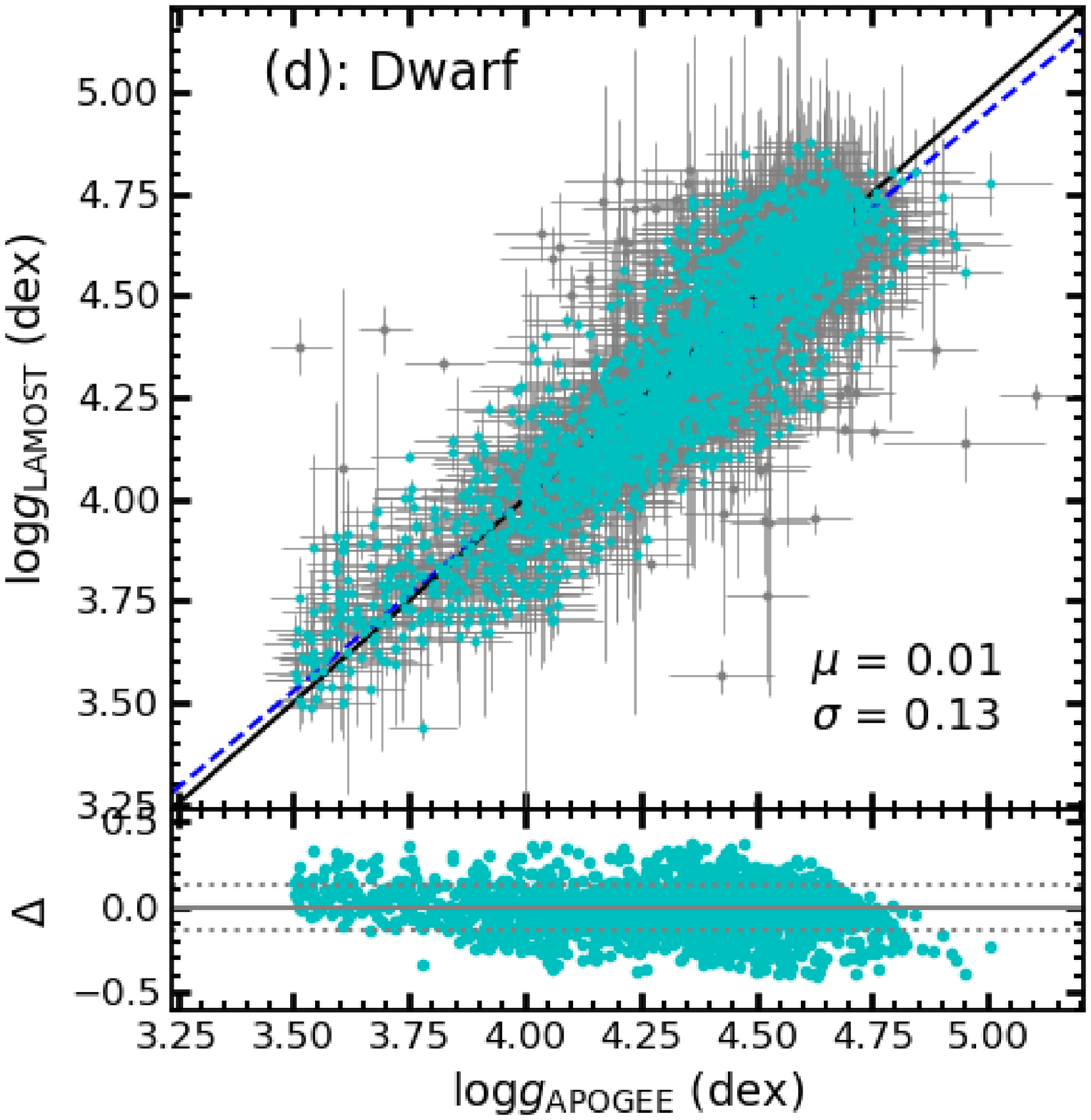}\includegraphics[width=6cm]{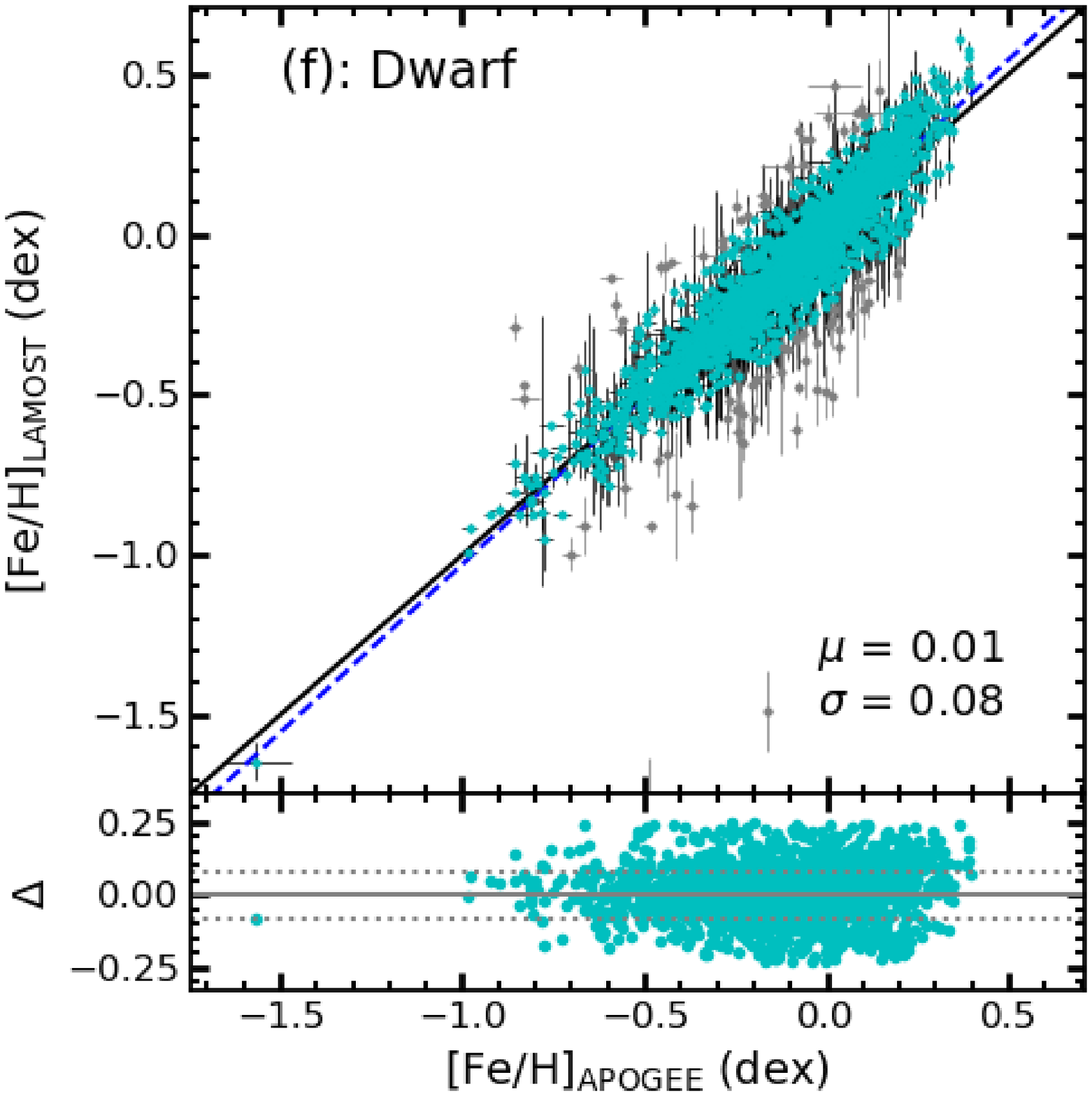}
\flushleft
\includegraphics[width=6cm]{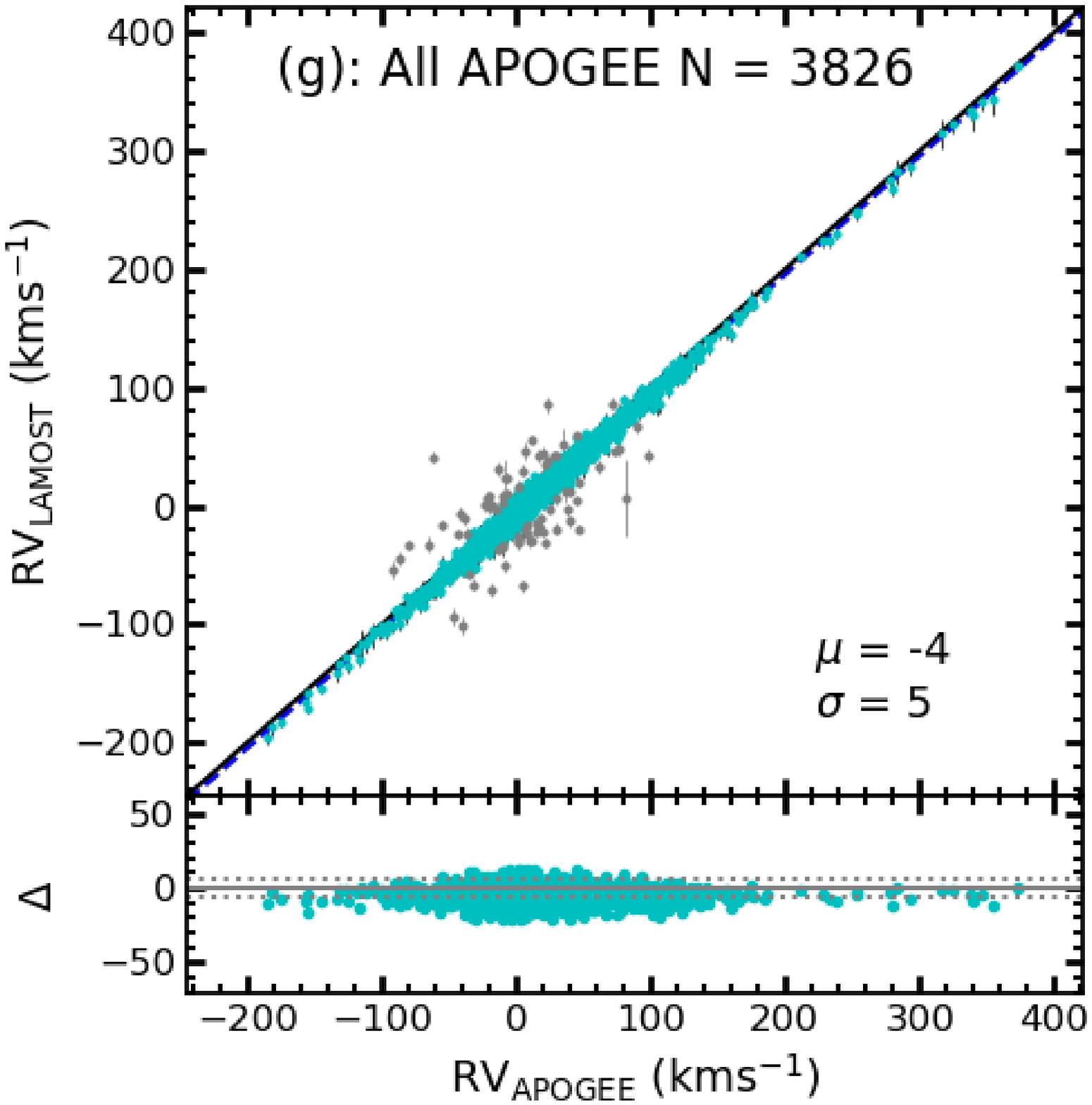}\includegraphics[width=6cm]{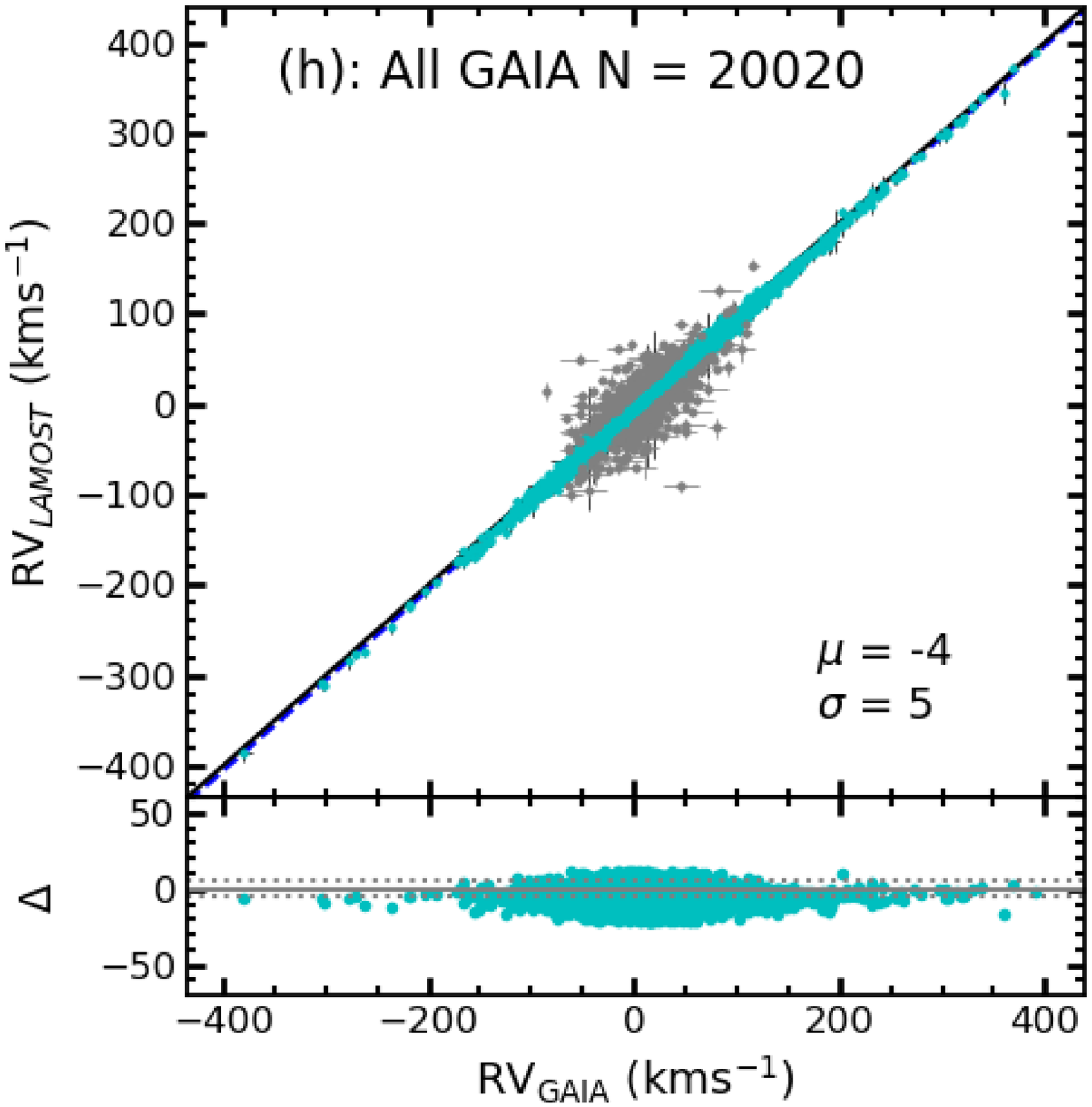}
\caption{
The top panels are the comparison of the atmospheric parameters of L\Ktwo\ giants with APOGEE, and the middle panels are the comparison of L\Ktwo\ dwarfs with APOGEE. The number of giants and dwarfs is 1,307 and 2,519, respectively. The bottom panels are the comparison of the RV of the L\Ktwo\ with APOGEE (left) and GAIA (right), respectively. The black solid and blue dashed lines indicate the bisector and optimal linear fitting lines, respectively. The grey dots indicate the targets whose values deviate more than 3$\sigma$ from the average, where the $\sigma$ level is shown by grey dashed lines in the bottom panels. \label{comp}
}
\end{figure*}

Figure\,\ref{comp} shows the results of the external comparisons with other data samples:
i.e. the comparison 
of atmospheric parameters and radial velocities between the L\Ktwo\ non-variable target and APOGEE
(panels a--g) and GAIA (panel h).
We clearly see that \teff\ agrees well between the different catalogs (panels a and b). Both the giants and dwarfs are located
around the bisector for \teff. The residuals, defined here as
\begin{equation}
\Delta P = P_{\rm LAMOST} - P_{\rm APOGEE/GAIA},
\end{equation}
are found with a bias value of $\mu = 2$ and 38\,K, and a standard deviation of $\sigma = 83$ and 118\,K for the giants and dwarfs, respectively.
A linear regression, expressed as \textbf{$y = a_2 x + b_2$}, was also applied to these plots, with the best-fit values listed in Table\,\ref{tab:excalib}.
The coefficients confirm the good agreement 
between the \teff\ values of L\Ktwo\ and APOGEE.
We note there are also a few outliers whose residuals deviate more than 3$\sigma$ from the mean level.
The comparison of \logg\ values is presented in 
Figs.\,\ref{comp}(c) and (d)
for giant and dwarf stars, respectively. The giants, compared to APOGEE, show consistent results
in general, as further indicated by the best-fit coefficient $a_2 = 0.95$ 
being very close to unity.
However, the \logg\ values from LAMOST are slightly higher than those from APOGEE, with a bias of $\mu =0.12$\,dex and a deviation of $\sigma = 0.16$\,dex.
We note that the scatter becomes larger for $\log g\,\leq 2.0$\,dex.
The \logg\ comparison for dwarf stars shows a slightly larger scatter. 
In addition, the fitting coefficient, $a_2 = 0.92$, is slightly lower than the one for the giant stars. This is evident from the slope of the residuals, which is smaller than the bisector.
The results for \feh\ are better than those for \logg.
We see a good agreement for \feh\ of giants (Fig.\,\ref{comp}(e)), confirmed by the best-fit coefficient $a_2 = 1.00$.
The \feh\ comparison for dwarfs displays a linear relation with a slope of $a_2 = 1.05$. 
All the linear fitting coefficients are provided in Table\,\ref{tab:excalib}.
Fig.\,\ref{comp}(g) and \ref{comp}(h) show the RV comparison for stars in common with the APOGEE and GAIA DR2 catalogs, respectively.

The following equation provides a linear regression for RV data:
\begin{eqnarray}
\left \{
\begin{array}{l}
{\rm RV}_{\rm LAMOST} = (1.00 \pm 0.01)\times {\rm RV}_{\rm APOGEE} - (5 \pm 1)\ {\rm km\,s}^{-1}, \\
{\rm RV}_{\rm LAMOST} = (1.00 \pm 0.01)\times {\rm RV}_{\rm GAIA} - (5 \pm 1)\ {\rm km\,s}^{-1}.
\end{array}
\right.
\end{eqnarray}
{\noindent 
It is clear that these best-fit lines are parallel to the bisectors but with a bias value of 5\,\kms. 
In general, the external comparison between L\Ktwo\ and APOGEE shows a good agreement for both the atmospheric parameters and RV.}

\startlongtable
\begin{deluxetable*}{llllllll}
\tablecaption{External calibration parameters.
\label{tab:excalib}}
\tablehead{
{} & \multicolumn{3}{c}{Giant} & {} & \multicolumn{3}{c}{Dwarf}\\
\colhead{} &
\colhead{\teff} &
\colhead{\logg} &
\colhead{\feh} &
\colhead{} &
\colhead{\teff} &
\colhead{\logg} &
\colhead{\feh}
}
\startdata
$a_2$         & 0.97 $\pm$ 0.01 & 0.95 $\pm$ 0.01 & 1.00 $\pm$ 0.01  &   & 1.13 $\pm$ 0.01 & 0.92 $\pm$ 0.01 & 1.05 $\pm$ 0.01 \\
$b_2$         & 147 $\pm$ 55   & 0.24 $\pm$ 0.02 & -0.02 $\pm$ 0.01 &   & -658 $\pm$ 28   & 0.35 $\pm$ 0.04 & 0.02 $\pm$ 0.01 \\
$\sigma_{ex}$ & 83           & 0.16          & 0.07           &   & 118          & 0.13      & 0.08  \\
$\mu_{ex}$    & 2            & 0.12          & -0.01          &   & 38           & 0.01      & 0.01  \\
\enddata
\end{deluxetable*}

\subsection{Calibration of LAMOST parameters}
\label{subsec:calibr}

In Sects.~\ref{sec:In-calibration} 
and \ref{sec:Ex-calibration}, we estimated the internal uncertainties of the atmospheric parameters and RVs derived from L\Ktwo\ spectra and compared the results to two external surveys APOGEE and GAIA. 
On the basis of those comparisons, here we put forward calibration relations that can be used to put the LAMOST parameters on the same scale
of APOGEE and GAIA data. Moreover, the linear regressions discussed in Section~\ref{sec:Ex-calibration} provide the associated uncertainties with the propagation errors,
from the following equations:
\begin{eqnarray}
\left\{
\begin{array}{l}
P_i = (P_{i,\rm{LAMOST}} - b_2)/a_2, \\
\sigma_i = \sqrt{\sigma_{\rm{in}}^2 + \sigma_{\rm{ex}}^2}.
\end{array}
\right.
\label{eq:calibration}
\end{eqnarray}
{\noindent where, as before, the index $i \in [1,N]$ indicates the $i$th measurement, $P_i$ denotes the calibrated parameters, $P_{i, \rm {\rm LAMOST}}$ represents the LAMOST parameters, and $a_2$ and $b_2$ represent the slope and the zero-point
of the linear regressions whose values are reported in Table\,\ref{tab:excalib}, $\sigma_{\rm{in}}$ is the unbiased internal error that can be calculated for each \snrg\ value
through Eq.\,(2), and $\sigma_{\rm{ex}}$ is the external deviation of the each parameter as shown in Figure\,\ref{comp}.}

\begin{figure*}[ht!]
\centering
\includegraphics[width=15cm]{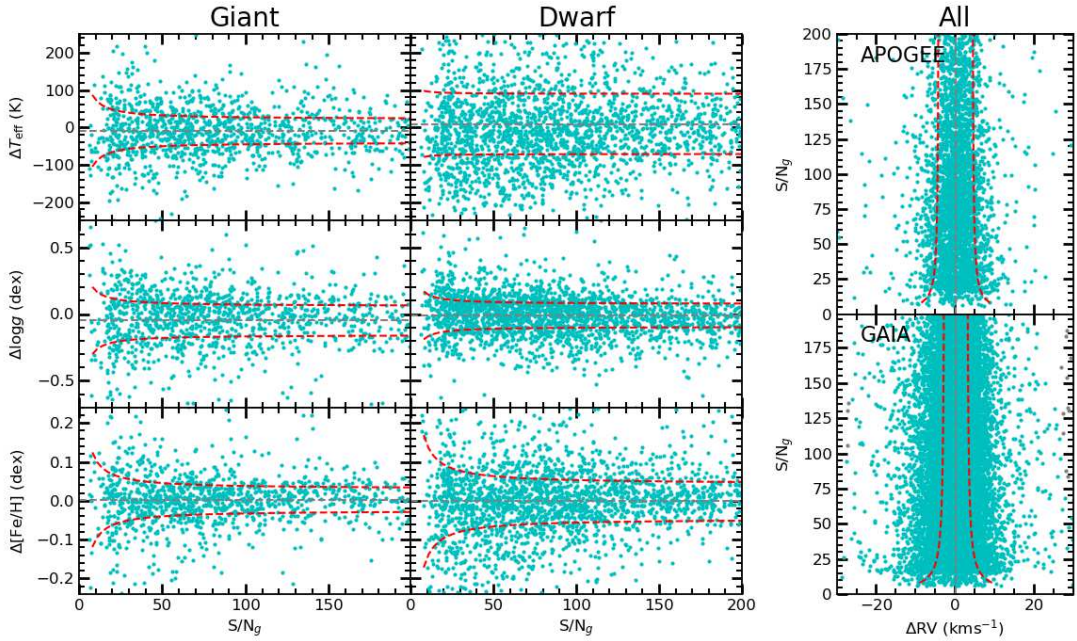}
\caption{
Distribution of the differences between LAMOST 
and literature parameters, $\Delta P$, after the application of both internal and external calibrations.
The left panels refer to giants and dwarf, while right panels are all targets' RVs for APOGEE and GAIA.
The mean values are all close to zero and are indicated by the horizontal lines. 
The red dashed curves in each box represent the 1$\sigma$ levels of the best fits as a function of the \snrg\ derived by means of Eq.\,\ref{eq:fit}. 
The gray points indicate that the outliers in the bin are beyond 6$\sigma$, and the cyan points are used for fitting.
\label{expe}}
\end{figure*}

The calibration is applied for each parameter independently 
for
two groups of stars, i.e., the giants and dwarfs, distinguished on the basis 
of their LAMOST \logg\ value,
as explained in Section~\ref{sec:Ex-calibration}. 
The left panels  
of Figure\,\ref{expe} show the distributions of the differences between the calibrated LAMOST values of the atmospheric parameters (\teff, \logg, and \feh) and their corresponding ones in the APOGEE catalog for the giants and dwarfs while the right panels show the same for the calibrated LAMOST RV values for the stars in common with the APOGEE (top) and GAIA DR2 (bottom) catalogs.
The mean values of these differences are very small for most of the derived physical quantities, which supports the good agreement between the two data sets after application of the calibrations (Eq.~\ref{eq:calibration}).
The dispersion of the final errors are fitted with a reciprocal relation of the form of Equation\,\ref{eq:fit} but with coefficients  $a_3$ and $b_3$. 
Their best-fit values are reported in Table\,\ref{tab:calib} along with $\mu$, referring to the mean values of the residuals.

\startlongtable
\begin{deluxetable*}{lrrrrrrrrr}
\tablecaption{Calibration parameters.\label{tab:calib}}
\tablehead{
{} & \multicolumn{4}{c}{Giant} & {} & \multicolumn{4}{c}{Dwarf}\\
\colhead{} &
\colhead{\teff} &
\colhead{\logg} &
\colhead{\feh} &
\colhead{RV} &
\colhead{} &
\colhead{\teff} &
\colhead{\logg} &
\colhead{\feh} &
\colhead{RV}
}
\startdata
$a_3$ & 519.77 & 1.15 & 0.75 & 35.71 &  & 64.71 & 0.76 & 1.00 & 51.31 \\
$b_3$ & 31.02  & 0.11 & 0.03 & 4.22  &  & 80.46 & 0.09 & 0.04 & 2.78 \\
$\mu$ & -9     & -0.05 & 0.00 & 0    &  & 9    & -0.09  &-0.04 & 0  \\
\enddata
\end{deluxetable*}

\begin{figure}[ht!]
\centering
\includegraphics[width=10cm]{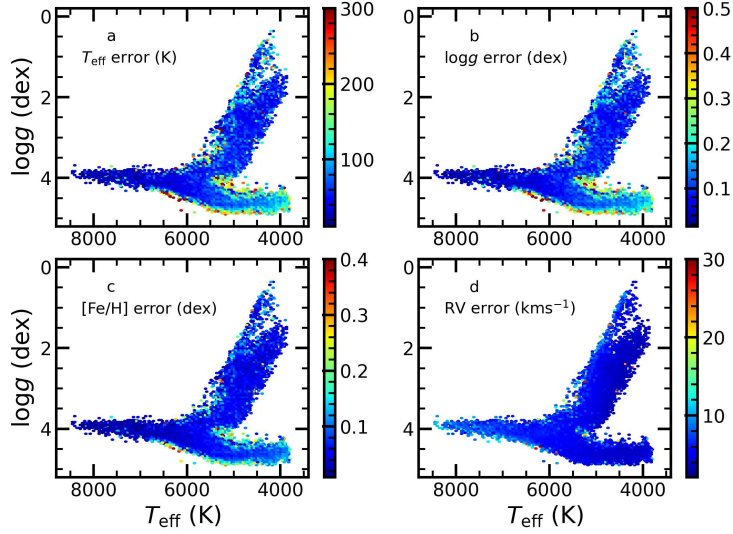}%\includegraphics[width=6cm]{fig-add01}
\caption{The Kiel (\teff\ v.s. \logg) diagram of the L\Ktwo\ samples. Note that the colors on the panels represent the calculated errors of atmospheric parameters (\teff, \logg, and \feh) and RV in each bin (see text for details). \label{kiel-e}}
\end{figure}

Figure \,\ref{kiel-e} shows the distributions of derived errors associated with atmospheric parameters across the Kiel diagram. The entire diagram are divided by a 100$\times$100 bin grid. We calculated the mean errors in each bin individually whose values are indicated by their colors. We can clearly see the errors of the stellar parameters derived with LASP are almost homogeneously distributed on the Kiel diagram, except a few values along the edge.

\section{Statistical analysis of stellar parameters} 
\label{sec:distri}

\begin{figure}[ht!]
\centering
\includegraphics[width=9cm]{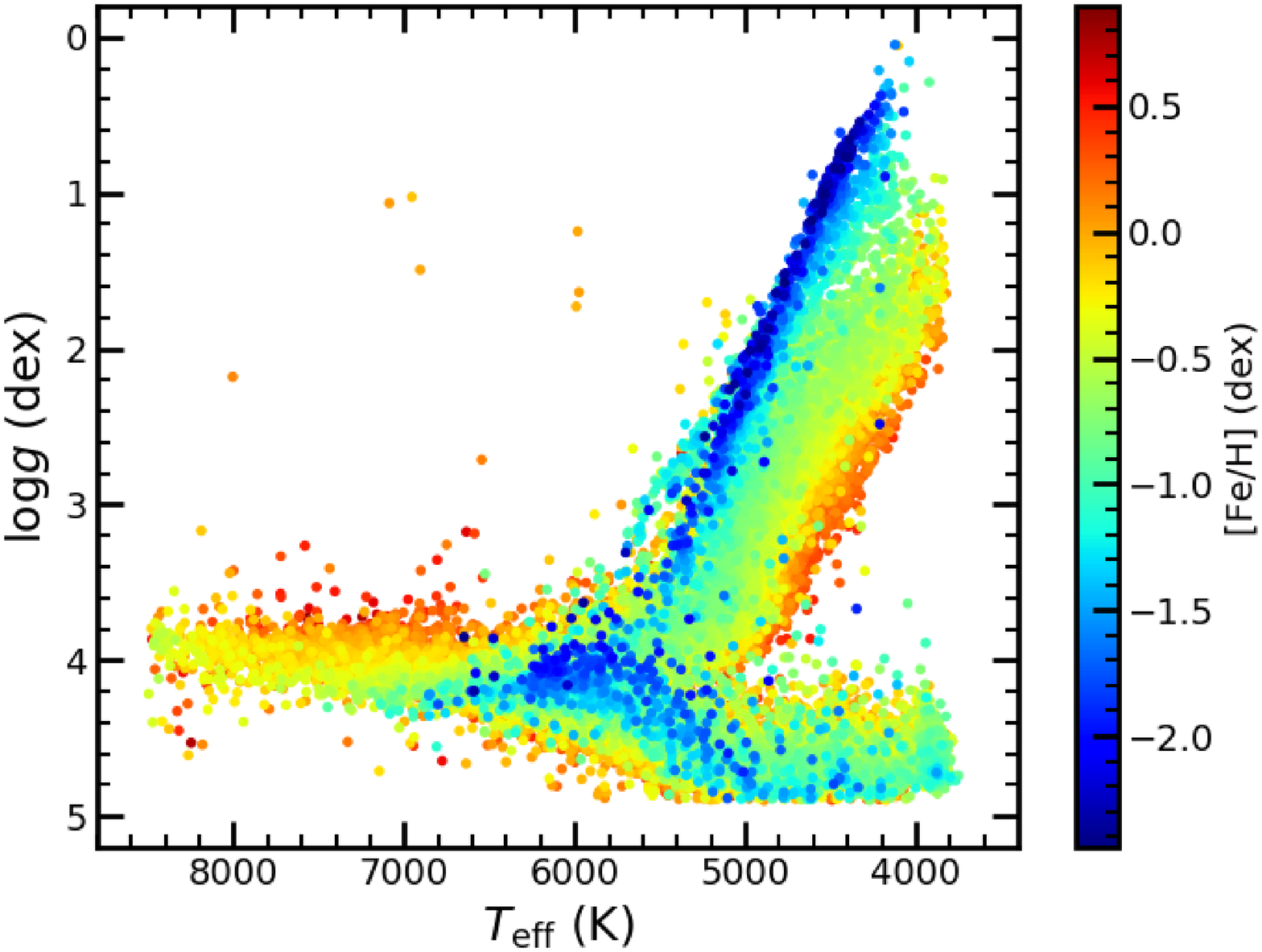}\includegraphics[width=9cm]{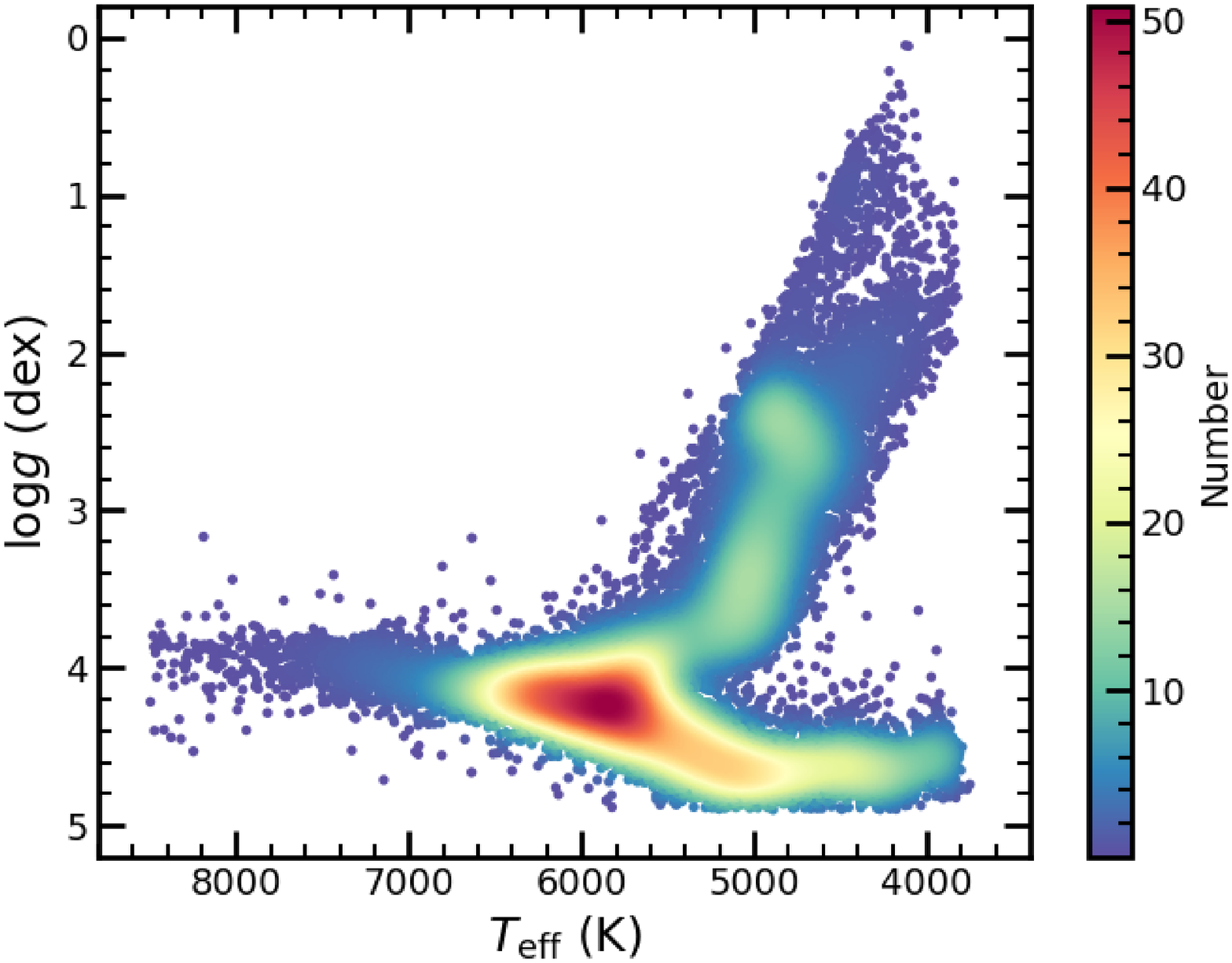}
\caption{The Kiel diagram of the LAMOST parameters derived from the 53,421 
qualified spectra of the L\Ktwo\ sample. Note that the colors on the left and right panels represent \feh\ and density number respectively.} 
\label{kiel}
\end{figure}

So far, the LAMOST-\Ktwo\ project has produced 160,619 low-resolution LAMOST spectra of 84,012 stars, including 70,895 objects with derived atmospheric parameters. More than 30,000 stars were observed at multiple epochs.
As shown in Figure\,\ref{err}, the internal uncertainties of the parameters decrease as their \snrg\ increases. For objects observed more than once, 
we adopt the parameters derived for the spectrum with the highest \snrg.
Figure\,\ref{kiel} shows the \logg\ - \teff\ plane (Kiel diagram) for the sources in the L\Ktwo\ sample. 
It is apparent how most of the objects are situated on the main sequence and the giant branch, 
among which the main sequence is indeed the longest phase in the lifetime of a star.

%%in which the main sequence is indeed the longest phases in the life-time of a star.
As the LASP pipeline works properly for AFGK-type stars only, the \teff\ values are found in the range from 3800\,K to 8400\,K. 
The values of \logg\ are found in the range of [0, 5.0]\,dex. 
The majority of the stars have an \feh\ value that is close to solar, but
low-metallicity stars are also present in our sample.
Similar to \citet{2018ApJS..238...30Z}, the giant branch correctly displaces toward higher temperatures as \feh\ decreases. 
We also note that a slight upward trend was found in the range of low temperature on the main sequence.

\begin{figure*}[ht!]
\centering
\includegraphics[width=17cm]{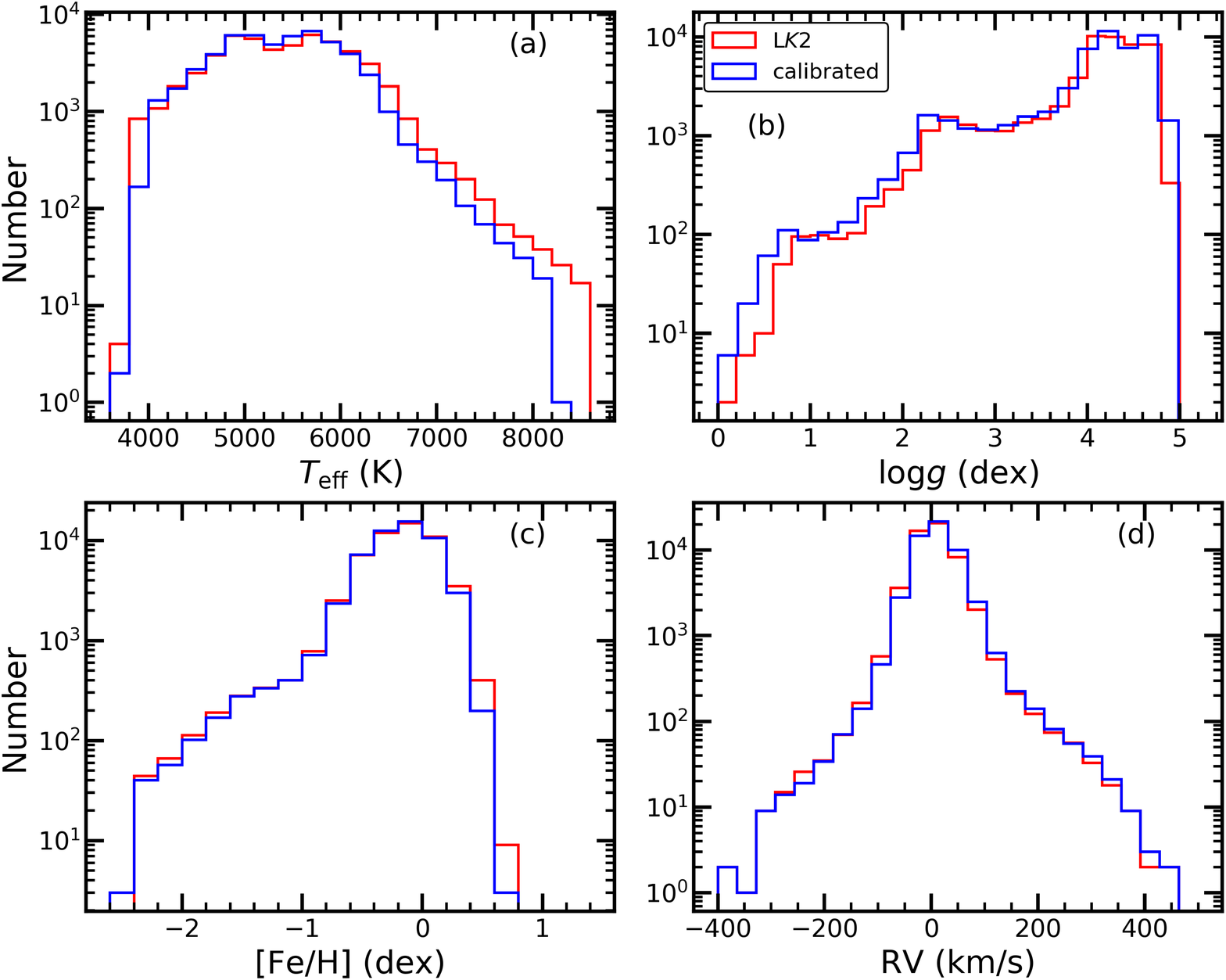}
\caption{Distributions of stellar parameters: \teff\ (a), \logg\ (b),  \feh\ (c), and RV (d). Original L\Ktwo\ and calibrated
data are represented with red and blue histograms, respectively, as indicated in the legend. \label{hist}}
\end{figure*}

The histograms plotted in Figure\,\ref{hist} shows the distributions of atmospheric parameters (\teff, \logg, and \feh) and radial velocity (RV) before and after calibration. 
\teff\ displays two peaks around 4800\,K and 5600\,K, respectively, in both data sets, which is likely the result of the projection on the \teff\ axis of the main-sequence and red giant branch.
The two \teff\ distributions are similar to each other, but the corrected one is slightly displaced towards cool temperatures. 
The histogram of \logg\ reveals a bimodal distribution with peaks around 2.4\,dex and 4.2\,dex, which, as before, could be the fingerprint of data clustering around the main sequence and giant branch.
However, the lower peak at 2.4 dex and 0.8 dex becomes slightly clearer after applying the calibration. 
This might be due to the correction made by the linear relation with a small slope that smooths out the peaks.

There is hardly any difference between the distributions of \feh\ before and after the calibration. Most stars have a close-to solar metallicity. 
The two distributions of RV, before and after calibration, are very similar. 
They are centered around 0\,\kms. There are few stars with $|RV|>300$\,\kms\ that can be classified as candidate high-velocity stars.
In general, the 
distributions of the stellar parameters derived from our sample are similar to those shown by
\citet{2015ApJS..220...19D} and \citet{2018ApJS..238...30Z}.

\section{Summary} \label{subsec:disc}
The \Ktwo\ mission has collected high-precision photometry for more than 400,000 stars with a time span of $\sim 80$\,days for each source.
These high-quality data pave the pathway to many different fields of astrophysics, such as asteroseismology, stellar activity and exoplanet research
\citep{2015ApJ...809L...3S, 2015ApJ...806..215F, 2018MNRAS.477.1120K}. Even though \cite{2016ApJS..224....2H} provide a catalog of stellar parameters for the objects in the first eight \Ktwo\ campaigns, they use different methods to deduce the values of parameters and from different instruments.
In the present work we report on the largest homogeneous spectroscopic dataset for the L\Ktwo\ sources that is based on LAMOST spectra.
Compared to the \kepler\ field \citep{2018ApJS..238...30Z}, the \Ktwo\ campaigns are better suited for observations
with LAMOST because the even distribution
of the \Ktwo\ fields on the ecliptic plane fits better with the observing constraints of LAMOST, with the exception of the \Ktwo\ fields with a very low declination (DEC $<\ -10\degr$).

The L\Ktwo\ project started in 2015 and has observed 126 plates across 15 \Ktwo\ campaigns up to 2018 February. 
Thanks to the wide distribution of \Ktwo\ fields in right ascension,
there are many objects in common with other LAMOST surveys.
After cross-matching the catalogs, we have collected a total of 160,619 spectra of 84,012 \Ktwo\ sources from LAMOST DR6. 
%The matching criterion is defined with a search radius of 3.7 arcsecs.
The cross-match is based on a search radius of at maximum 3.7 arcsecs around the observed equatorial coordinates.
However, the angular
separation of 94.36\% of the sources 
in the two catalogs is less than 1.0 arcsec (Figure\,\ref{fig:cross}).
Our catalog now covers {20.68\,\%} \Ktwo\ objects spread over all the \Ktwo\ campaigns observable with LAMOST.
As LAMOST can only point to targets with declination higher than $-10$ degrees, the fraction of observed targets increases to 27.38\,\% of the all the \Ktwo\ objects observable with LAMOST.
The atmospheric parameters and the radial velocities provided in this paper have been derived through the LASP pipeline for 80.92\% of the L\Ktwo\ spectra, covering 70,895 individual \Ktwo\ targets. We note, however, that the LASP works only for stars of the A, F, G, or K spectral type, and does not deliver atmospheric parameters for the O, B, and M-type stars. That is unfortunate because the latter stars are subjects of many different types of research, ranging from the search of exoplanets in the habitable zones of M-type dwarfs \citep{2017Natur.542..456G} through the study of the internal structure of pulsating hot OB sub-dwarfs. Since those investigations rely heavily on the stars' atmospheric parameters, it is very important to derive their values also for those stars observed in the framework of the L\Ktwo\ projects which fall outside the limits of the LASP using other methods \citep[see, e.g.,][]{2019ApJ...881..135L, 2019ApJ...881....7L}. That is, however, beyond the scope of this paper.

We estimated the internal uncertainties for the \teff, \logg, \feh, and RV through the 
results obtained for
objects with  multiple visits. 
We found average uncertainties of {28\,K, 0.05\,dex, 0.03\,dex and 3\,\kms} for \teff, \logg, \feh, and RV at \snrg\,$\sim 50$, respectively. 
The precision improves as \snrg\ increases. 
This result is half that of \citet{2016ApJS..225...28R},
%in which they found that the precision is
who found a precision of
 68\,K, 0.08\,dex and 0.06\,dex for \teff, \logg, and \feh\ at \snrg\,$\sim 50$.
The external accuracies of the stellar parameters of the targets of the
L\Ktwo\ sample are evaluated by comparison 
to APOGEE and GAIA DR2 catalogs, respectively.
We found that, in general, our stellar parameters for giant and dwarf stars agree well with those provided by the APOGEE survey
as their values are closely following a one-to-one relation.

This is possibly the result of  the large errors of the dwarf values of \logg\ and \feh\ in common with L\Ktwo\ compared to APOGEE.
These fitting slopes are likely responsible for the differences of the distributions of \teff\ and \logg\ before and after correction (see Figure\,\ref{expe}).

In addition to the LASP pipeline, there are other codes that have been applied on the L\Ktwo\ spectra, such as MKCLASS \citep{2014AJ....147...80G} and ROTFIT \citep{2016A&A...594A..39F}, to derive spectral types, stellar parameters and RVs. The comparison between ROTFIT and LASP {\bf (v.2.7.5)} 
showed that the results of both methods are in general consistent with each other \citep{2016A&A...594A..39F}. As the low-resolution spectra cover almost all the visible wavelengths, they can also be used to calculate 
indexes of stellar activity from the equivalent width of Ca\,{\sc ii}\,H and K lines \citep[see, e.g.,][]{2008AJ....135..785W,Karoffetal2016}, Ca\,{\sc ii}\,IRT, and H$\alpha$ \citep[see, e.g.,][]{2016A&A...594A..39F}.

We recall that the L\Ktwo\ program will continue in the next few years. All the spectra will be publicly available 
from 2021 onwards \footnote{http://dr6.lamost.org/}.
The experiences gained from this project inspired us to initiate phase II of the LAMOST-{\sl Kepler}/\Ktwo\ survey, which is a new parallel 
program to observe time series of medium-resolution LAMOST spectra for a selection of 20 footprints distributed over the {\sl Kepler} field and \Ktwo\ campaigns \citep{2020arXiv200906843Z}.
For the stars in common, the LAMOST observations with two different spectral resolutions can be analysed simultaneously for a better understanding of these sources and to study other astrophysical phenomena, such as the orbits of binaries and the short-term evolution of stellar activity.
We foresee a wide usage of 
the spectra of the L\Ktwo\ project in the near future.

\acknowledgments
We acknowledge support from the Beijing Natural Science Foundation (No. 1194023) and the National Natural Science Foundation of China (NSFC) through grants 11673003, 11833002 and 11903005. 
WZ is supported by the Fundamental Research Funds for the Central Universities.
The Guoshoujing Telescope (the Large Sky Area Multi-object Fiber Spectroscopic Telescope LAMOST) is a National Major Scientific Project built by the Chinese Academy of Sciences. Funding for the project has been provided by the National Development and Reform Commission.
LAMOST is operated and managed by the National Astronomical Observatories, Chinese Academy of Sciences. JTW, JNF and WKZ acknowledge the support from the Cultivation Project for LAMOST Scientific Payoff and Research Achievement of CAMS-CAS. 
WKZ and JXW hold the LAMOST fellowship as a Youth Researcher which is supported by the Special Funding for Advanced Users, budgeted and administrated by the Center for Astronomical Mega-Science, Chinese Academy of Sciences (CAMS). 
This paper is dedicated to the 60th anniversary of the Department of Astronomy of Beijing Normal University, the 2nd one in the modern astronomy history of China. JM-\.Z acknowledges the Wroclaw Centre for Networking and Supercomputing grant no.224. 
M.C.S acknowledge support from the National Key Basic Research and Development Program of China (No. 2018YFA0404501) and NSFC grant 11673083.
The work presented in this paper is supported by the project "LAMOST Observations in the {\it Kepler} field" (LOK), approved by the Belgian Federal Science Policy Office (BELSPO, Govt. of Belgium; BL/33/FWI20).

%% This command is needed to show the entire author+affilation list when
%% the collaboration and author truncation commands are used.  It has to
%% go at the end of the manuscript.
%\allauthors

%% Include this line if you are using the \added, \replaced, \deleted
%% commands to see a summary list of all changes at the end of the article.
%\listofchanges

\end{document}

%% file: Table02.tex
801037 & 211203556 & 03:59:23.34 & 26:32:04.97 & 1.77 & 15.40 & M1 & 2011\-10\-27T18:13:00 & 4 & 0.00 & spec\-55862\-B6210\_sp01\-037.fits \\
801230 & 211189643 & 03:58:28.36 & 26:12:55.61 & 2.02 & 16.14 & M0 & 2011\-10\-27T18:13:00 & 4 & 0.00 & spec\-55862\-B6210\_sp01\-230.fits \\
901054 & 202081906 & 06:28:37.96 & 26:23:26.87 & 3.12 & 15.70 & G8 & 2011\-10\-27T20:17:00 & 0 & 0.00 & spec\-55862\-B6212\_sp01\-054.fits \\
... &  &  &  &  &  &  &  &  &  &  \\

%% file: Table03.tex
407004121 & 201176436 & 11:30:41.60 & \-04:39:42.00 & 4250 $\pm$ 55 & 4.73 $\pm$ 0.09 & \-0.53 $\pm$ 0.05 & \-22 $\pm$ 4 & -\\
558704121 & 201176436 & 11:30:41.60 & \-04:39:42.00 & 4232 $\pm$ 56 & 4.71 $\pm$ 0.09 & \-0.51 $\pm$ 0.05 & \-20 $\pm$ 4 & -\\
499814067 & 201238068 & 11:36:19.20 & \-03:23:24.00 & 4170 $\pm$ 55 & 4.67 $\pm$ 0.09 & \-0.33 $\pm$ 0.05 & 7 $\pm$ 4 & -\\
... & ... & ... & ... & ... & ... & ... & ... & ... \\